\documentclass[12pt]{article}

\makeatletter
\@addtoreset{equation}{section}

\makeatother

\begin{document}
\setlength{\unitlength}{0.2cm}

\title{
Universality of subleading corrections for self-avoiding
walks in presence of one-dimensional defects
}
\author{
  \\
  {\small Sergio Caracciolo }             \\[-0.2cm]
  {\small\it Scuola Normale Superiore and INFN -- Sezione di Pisa}  \\[-0.2cm]
  {\small\it I-56100 Pisa, ITALIA}          \\[-0.2cm]
  {\small e-mail: {\tt Sergio.Caracciolo@sns.it}}     \\[-0.2cm]
  \\[-0.1cm]  \and
  {\small Maria Serena Causo}             \\[-0.2cm]
  {\small\it Dipartimento di Fisica and INFN -- Sezione di Lecce}  \\[-0.2cm]
  {\small\it Universit\`a degli Studi di Lecce}        \\[-0.2cm]
  {\small\it I-73100 Lecce, ITALIA}          \\[-0.2cm]
  {\small e-mail: {\tt causo@le.infn.it}}     \\[-0.2cm]
  \\[-0.1cm]  \and
  {\small Andrea Pelissetto}                          \\[-0.2cm]
  {\small\it Dipartimento di Fisica and INFN -- Sezione di Pisa}    \\[-0.2cm]
  {\small\it Universit\`a degli Studi di Pisa}        \\[-0.2cm]
  {\small\it I-56100 Pisa , ITALIA}          \\[-0.2cm]
  {\small e-mail: {\tt pelisset@ibmth.difi.unipi.it}}   \\[-0.2cm]
  {\protect\makebox[5in]{\quad}}  
  \\
}
\vspace{0.5cm}

\maketitle
\thispagestyle{empty}   

\vspace{0.2cm}

\begin{abstract}
We study three-dimensional self-avoiding walks in presence of 
a one-dimensional excluded region. We show the appearance of a 
universal sub-leading exponent which is independent of the particular
shape and symmetries of the excluded region. A classical 
argument provides the estimate: $\Delta = 2 \nu - 1 \approx 0.175(1)$. 
The numerical simulation gives $\Delta = 0.18(2)$.
\end{abstract}

\clearpage

\newcommand{\be}{\begin{equation}}
\newcommand{\ee}{\end{equation}}
\newcommand{\bea}{\begin{eqnarray}}
\newcommand{\eea}{\end{eqnarray}}
\newcommand{\<}{\langle}
\renewcommand{\>}{\rangle}

\def\spose#1{\hbox to 0pt{#1\hss}}
\def\ltapprox{\mathrel{\spose{\lower 3pt\hbox{$\mathchar"218$}}
 \raise 2.0pt\hbox{$\mathchar"13C$}}}
\def\gtapprox{\mathrel{\spose{\lower 3pt\hbox{$\mathchar"218$}}
 \raise 2.0pt\hbox{$\mathchar"13E$}}}

\newcommand{\R}{\hbox{{\rm I}\kern-.2em\hbox{\rm R}}}

\newcommand{\reff}[1]{(\ref{#1})}

\section{Introduction}

An important problem in statistical mechanics is the study of the 
critical behaviour of systems in geometries with boundaries. One 
usually considers situations in which the geometric constraint 
changes the critical behaviour: in the renormalization-group 
terminology these are the cases 
in which the boundary is a {\em relevant} perturbation.
However there are also cases, as the one we are concerned with
in this	paper,
in which the leading critical behaviour is unchanged and the presence 
of the boundary appears as an {\em irrelevant} perturbation.

Maybe because of their name the role of {\em irrelevant}
operators is generally not well	studied. Indeed, by definition,
they do not modify the fixed point of the
renormalization-group transformations. In particular they do not
change the values of the critical exponents and manifest
themselves only through subleading corrections to the fixed-point
hamiltonian. Nonetheless in any actual computation the fixed-point
hamiltonian is replaced by an effective hamiltonian whose parameters
are tuned closed to criticality. This means that as soon as a
precise determination of the universal scaling behaviour is needed,
it becomes important to have good control also on the terms
responsible for corrections to scaling.

In this paper, we will concentrate on three-dimensional 
self-avoiding walks
(SAWs) in presence of an excluded one-dimensional region. We will
extend the results of \cite{Considine-Redner,CFP,Grassberger} which showed 
for the case of a half-line the appearance of a new critical exponent
$\Delta\approx 0.22$. Here we will consider more general
one-dimensional regions and we will show that the value of $\Delta$
is independent of the shape and symmetries of the excluded region:
only the dimensionality plays a role. Moreover we will show that 
$\Delta$ can be predicted to a very good accuracy by a purely 
geometrical argument. 

The paper is organized as follows: in Section 2 we give a general discussion
of the relevance of the perturbation introduced by the excluded region
and we give a prediction for the subleading exponent. In Section 3 we describe 
the geometries we analyze and the choice of observables which allow the 
easiest and most precise determination of $\Delta$. In Section 4 we give a 
few details of the simulation, while in Section 5 we give the final results.
Appendix A presents the computation of the generating function 
for ordinary random walks in presence of excluded hyperplanes 
which allow a direct check of the results of Section 2. 
An analogous computation for the case a single excluded point
and for an excluded needle is reported in \cite{Considine-Redner}.  Appendix 
B contains some unrelated results on the small-momentum behaviour of the 
two-point function.

\section{Excluded set and corrections to scaling}

In the terminology of the field-theoretic approach to critical phenomena,
the critical behaviour of the SAW is governed by the fixed-point of the $O(n)$
$\sigma$-model analytically continued to 
$n=0$~\cite{Daoud,DC,Emery,ACF,FFSbook,DC-J}.
The presence of an excluded region corresponds to a perturbation
due to the introduction of an operator which creates vacancies in the
$O(n)$ model.
Now consider
the correlation function\footnote{In the SAW language we have
\be
G_{\cal R}(\vec{r};\beta) = \sum_{N=0}^\infty \beta^N c_{N,{\cal R}}(\vec{r})
\ee
where $c_{N,{\cal R}}(\vec{r})$ is the number of SAWs going from 0 to $\vec{r}$
in $N$ steps without intersecting the region ${\cal R}$.}
between a spin at the origin and one in the bulk
at location $\vec{r}$;  this function will have the scaling form
\be
G_{\cal R}(\vec{r};\beta)  \;\sim\;
   r^{- (d-2+\eta_{\cal R})} F_{\cal R}(\vec{r}/ \xi(\beta)) \,+\,
   r^{- (d-2+\eta'_{\cal R})} F'_{\cal R}(\vec{r}/ \xi(\beta)) \,+\,
      \ldots   \;.
 \label{GI}
\ee
Here $\beta$ is the inverse temperature, and
$\xi \sim (\beta_c-\beta)^{-\nu}$ is the correlation length
{\em in the unperturbed theory}\/;
the critical inverse temperature $\beta_c$ and
the exponent $\nu$ are {\em not}\/ modified
by the presence of the vacancies,
unless the excluded region ${\cal R}$ is so big
that the remaining set ${\bf Z}^d \setminus {\cal R}$ is
effectively a space of lower dimensionality.
See Ref.~\cite{Stu} for how thin a set has to be before 
$\mu = 1/\beta_c$ changes.
However, the behaviour of the other quantities depends on whether the
perturbation is relevant or irrelevant \cite{Wegner}:
\begin{itemize}
\item[(a)]
  If the perturbation is {\em relevant}\/, then the leading spin-spin decay
  exponent $\eta_{\cal R}$ {\em differs}\/ from its bulk value $\eta$
  [and as a consequence the leading susceptibility exponent
  $\gamma_{\cal R} = (2-\eta_{\cal R}) \nu$
  differs from its bulk value $\gamma = (2-\eta) \nu$].
  Likewise, the leading scaling function $F_{\cal R}$
  differs from its bulk value $F$;
  in particular, it has a non-trivial angular dependence~\cite{Cardy-Redner}.
\item[(b)]
  If the perturbation is {\em irrelevant}\/, then
  $\eta_{\cal R}$ and $F_{\cal R}$ are {\em unchanged}\/ from their
  bulk values $\eta$ and $F$.
  In particular, the leading scaling function
  $F$ has {\em no angular dependence}\/.
  The effects of the perturbation show up only in the {\em non-leading}\/
  exponents and scaling functions $\eta'_{\cal R}, \ldots$
  and $F'_{\cal R}, \ldots$,
  which {\em can}\/ differ from their bulk values.
\end{itemize}
In either case, $\eta_{\cal R}$ and $F_{\cal R}$
(and indeed all of the exponents $\eta'_{\cal R}, \ldots$
and scaling functions $F'_{\cal R}, \ldots$ except for an unknown amplitude)
are universal in the sense
that they depend only on the {\em global}\/ properties of the excluded region
${\cal R}$, such as its dimensionality.

More subtle is the case in which the perturbation is marginal:
$\eta_{\cal R}$ is equal to the bulk value but the
universal scaling behaviour may be broken by logarithmic violations and
observables associated to the perturbation can show a complete breaking of
universality, in the sense that they can have critical exponent 
with an explicit
dependence from the coupling of the perturbation~\cite{Bariev,Eisen}.

To understand the effect of the introduction of the excluded region
we must thus understand if the perturbation is relevant or irrelevant.
We will resort to a geometric argument. Consider in a $d$-dimensional
space a set $E$ and 
let $d(E)$ denote the number of dimensions in which the set $E$
extends to infinity. Let us now
recall the fundamental rule in geometric probability for the
dimension of the 
generic intersection
$A\cap B$ of two geometric sets $A$ and $B$ which are
immersed in a space of $d$ dimensions:
\be
d(A\cap B) = d(A) + d(B) - d \label{dim} \;\; .
\ee
A negative sign of $d(A\cap B)$ means that $A$ and $B$
do not generically intersect in dimension $d$ outside any bounded volume.
The application of (\ref{dim}) extends also to random geometries where
one considers a probability measure on a configuration space
which is concentrated on a set of events with given Hausdorff
dimension. 

For example, the generic intersection of two
ordinary random walks, which have Hausdorff dimension 2, has
dimension $4-d$, and thus they do not generically intersect in
$d>4$. By using the well-known random-walk representation of the
euclidean $O(n)$-vector field theory~\cite{Symanzik-V,Jurg,FFSbook}, for
which the interaction is concentrated on the intersections among
walks, it is then possible to understand why for $d>4$ in the
critical region only a trivial theory is recovered: simply because
the walks intersect almost nowhere! We can say that for $d>4$:
\begin{itemize}
\item[a)] the probability of intersections of two random walks in
the critical region scales towards its limiting value with the
correlation-length as $\xi^{4-d}$;
\item[b)] critical indices take their free-field (that is
mean-field) values, the interaction term in the hamiltonian is {\em
irrelevant} and induces only sub-leading corrections.
\end{itemize}

In $d=4$ the interaction becomes {\em marginal} and is responsible
only for logarithmic corrections to the critical indices~\cite{ACF}.
It is said that 4 is the upper critical dimension of the model.

In $d<4$ the interaction is {\em relevant} and thus it changes the critical
indices.

Similar ideas have been used to discuss the critical behaviour of
gauge theories and random surfaces, see for example~\cite{Parisi79}.

Let us now consider the case of walks in presence of an excluded 
region $\cal R$ of dimension $d_{\cal R}$. Consider first
ordinary random walks whose
Hausdorff dimension is two. Then the dimension of a generic 
intersection with the region $\cal R$ of dimension $d_{\cal R}$ is
\be
d_{int} = 2 + d_{\cal R} - d
\label{eq2.4}
\ee
The previous argument suggests the following cases
\begin{itemize}
\item $d_{\cal R}=0$:  this is the case in which we are excluding
only a finite set of lattice sites. The upper critical dimension
(with its logarithmic corrections) is
$d=2$; 
\item $d_{\cal R}=1$:   this is the case in which we are excluding
a finite set of one-dimensional lines. The upper critical dimension
(with its logarithmic corrections) is $d=3$. 
\end{itemize}

Formula \reff{eq2.4} can also be obtained from other considerations
\cite{Considine-Redner}.
If $\cal R$ is a $d_{\cal R}$-dimensional hyperplane, consider the projection
of the walk on ${\cal R}'$, the orthogonal complement of $\cal R$. 
Also the projection is a random walk. 
The probability of intersection will be then
the probability of first return to the point ${\cal R} \cap {\cal R}'$.
But the probability that a $\overline d$-dimensional random walk eventually 
passes through a point scales as $1/N^{\overline{d}/2 - 1}$ if 
$\overline{d}>2$, while it is one if $\overline{d}\le 2$. Since in our
case $\overline{d}= d - d_{\cal R}$, we see that for $d_{int}\ge 0$
the walk generically intersects the region $\cal R$, while for 
$d_{int} < 0$ the probability vanishes as 
$1/N^{-d_{int}/2}$, in agreement with our argument.
Finally notice that when $d_{\cal R} = d - 1$, one can apply 
the results for the so-called vicious walkers. Indeed the 
behaviour of $d$ one-dimensional vicious walkers\footnote{
Related work on vicious walkers can be found in 
Ref.~\cite{vicious,vicious2}.}
can be viewed as the behaviour of a $d$-dimensional walker 
in a space bounded by $d(d-1)/2$ $(d-1)$-dimensional 
hyperplanes~\cite{Fisher,Fisher-Huse}.

Let us now arrive at the case of SAW. Their Hausdorff dimension
is $1/\nu$ (see for example~\cite[Appendix B]{ACF}). For the
dimension of a generic 
intersection with the region $\cal R$ we get
\be
d_{int} = {1\over \nu} + d_{\cal R} - d
\ee
Let us consider once more a series of cases, remembering that for
$d\leq 4$  according to the Flory formula
$\nu\approx 3/(d+2)$, which is exact in $d=1,2$, must be corrected
by logarithmic violations in $d=4$, and is a good approximation in
$d=3$.
\begin{itemize}
\item $d_{\cal R}=0$:  Excluded points are a relevant perturbation
only if $d\leq 1/\nu$, that is in $d=1$. We remark that the
sub-case in which the excluded set consists of a single point
$\cal R = P$, when $\cal P$ is chosen to be a nearest-neighbour of
the origin of the walk, can be mapped into a problem {\em without}
vacancies. Indeed each walk of $N$ steps, starting from the origin,
can be seen as a walk of $N+1$ steps starting from $\cal P$ whose
first step is the previous origin. Under this mapping the asymmetry
induced by the exclusion of $\cal P$ can be seen as the
correlation between the position of the end-point of the walk 
with the direction of the first step.
\item $d_{\cal R}=1$:  in $d=1,2$ the perturbation is relevant, but
it is already irrelevant in $d=3$.
\end{itemize}
Thus in $d=3$ with a finite set $\cal R$ of excluded lines, we expect
that, regardless of their disposition is space, the probability that
a  walk intersects the region $\cal R$ scales
as $\xi^{2-1/\nu}$, or, since $\xi\sim N^{\nu}$, as $N^{2\nu -1}$.

In the field-theoretic language this dimensional argument implies in \reff{GI}
\be
\eta' = \eta - d_{int} = \eta + 2 - {1\over \nu} \;\; .
\label{etaprime}
\ee
Consequently, if $P_k(x)$ is a homogeneous polynomial of degree $k$ in 
the coordinates of the end-point of the walk, we have
\be
\< P_k(x) \>_{N,{\cal R}} = N^{k \nu} \left( A(k) + {B_{{\cal R}}(k)\over
N^\Delta } + \cdots \right) \label{coordinate}
\ee
where $\<\ \cdot\ \>_{N,{\cal R}}$ is the average in the ensemble of walks
of length $N$ that do not intersect $\cal R$ and $\Delta$ a subleading 
exponent. If \reff{etaprime} holds, we have
\be
\Delta = -d_{int}\nu = 2 \nu -1 \;\; .
\label{Deltapredicted}
\ee
In general we expect renormalization effects to change \reff{etaprime}
and thus \reff{Deltapredicted} introducing an anomalous dimension.
However experience with three-dimensional models indicates 
that these corrections, if not vanishing, are extremely small and 
thus we expect \reff{Deltapredicted} to be in any case a very good 
approximation. 

The amplitude $A(k)$ is not universal, because of an unknown scale factor,
but it should not depend on the excluded region. Moreover, since 
in the critical limit the distribution of the end-point is rotationally 
symmetric, $A(k)$ vanishes whenever $\int d \Omega_x P_k(x) = 0$
where $d \Omega_x$ is the normalized measure on the sphere 
$S^{d-1}$.
This suggests a very convenient way to compute the subleading exponent
$\Delta$
induced by the introduction of the excluded region. The idea is to consider 
observables which have zero expectation value in the rotationally-invariant
continuum limit and which do not vanish under the residual discrete 
symmetry that the lattice has after the introduction of the excluded region. 
For these quantities $A(k)=0$ so that the leading term scales as 
$N^{k\nu-\Delta}$ which makes the determination of $\Delta$ much easier.
Let us notice that in all this discussion we have always assumed 
that the first subleading exponent is related to the introduction of the 
excluded region. This is true for those quantities for which 
$\< P_k(x)\>$  would vanish in absence of the excluded region, i.e.
for those $P_k(x)$  which are not only non-rotationally invariant
but which are also not symmetric under the transformations of the 
cubic group. If instead $\< P_k(x)\>$ would not vanish even in absence 
of the excluded region, it is not clear a priori which exponent 
should show up first. For cubic-symmetric $P_k(x)$ for which 
$A(k)=0$ an extensive analysis  \cite{noieGuttmann,CPRV}
shows that in absence of any excluded region 
\be 
\< P_k(x)\> \sim N^{k \nu - \Delta_{latt}} 
\ee
with $\Delta_{latt} \approx 2 \nu$. For $d=3$ this exponent is much larger 
than \reff{Deltapredicted} and thus, in presence of an excluded region,
we expect all non-rotationally invariant $P_{k}(x)$ to behave 
as $N^{k\nu-\Delta}$ with $\Delta$ given by \reff{Deltapredicted}.
We should also mention that the exponent $\Delta$ defined here is 
unrelated to the exponent $\omega\nu\approx 0.5$ which controls the 
leading confluent correction for SAWs in absence of any excluded 
region and which has been the object of much attention in the 
literature~\cite{Baker,RG-francesi0,Majid-et-al,Sokal2}.

\section{The models}

We have  concentrated upon the problem of $3$--$d$ SAWs starting from the 
origin in presence of an excluded region $\cal R$
which consists of a finite collection of half-lines
--- we will call them ``{\em needles}" ---
along the coordinate axes. 
 
For this purpose we have studied the following cases (see figure~\ref{picture})
for the
region $\cal R$ (for a reason to be clarified later all the
needles start from a site whose distance from the origin is two:
for example the needle along the
positive $z$-axis will start from the site with coordinate $(0,0,2)$):
\begin{itemize}
\item[1)] a needle along one direction;
\item[2)] two needles along the positive and negative directions of an
axis; 
\item[3)] two needles along two different axes;
\item[4)] four needles in a plane along the two directions of two
axes;
\item[5)] three needles along three different axes;
\item[6)] six needles along the two directions of all the
three axes.
\end{itemize}
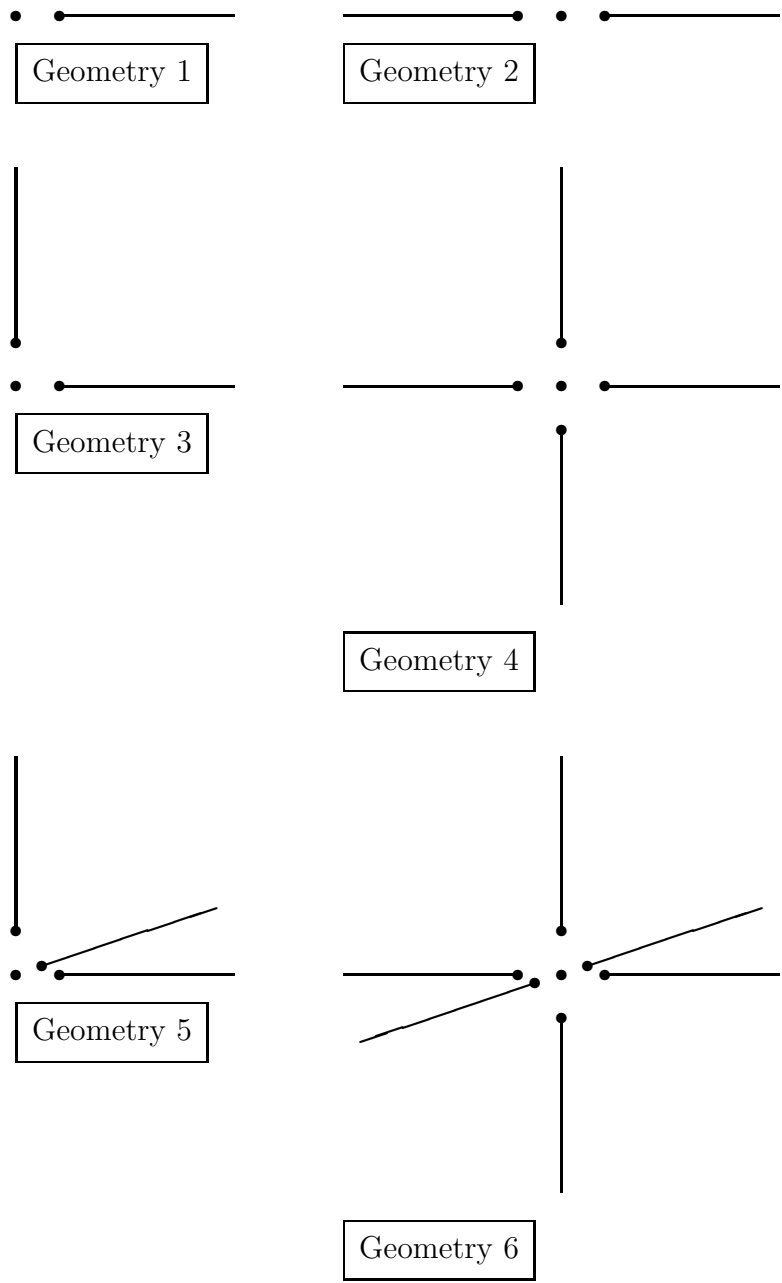
\begin{figure}[p]
\setlength{\unitlength}{0.29cm}
\begin{picture}(45,55)(-7,-60)
\thicklines
\setlength{\fboxsep}{0.2cm}
\put(-10,-3){
     \begin{picture}(20,4)(-10,-3)
     \put(0,0){\circle*{0.5}}
     \put(2,0){\circle*{0.5}}
     \put(2,0){\line(1,0){8}}
     \put(0,-3){\fbox{Geometry 1}}
     \end{picture}
     }
\put(15,-3){
           \begin{picture}(20,4)(-10,-3)
           \put(0,0){\circle*{0.5}}
           \put(2,0){\circle*{0.5}}
           \put(-2,0){\circle*{0.5}}
           \put(2,0){\line(1,0){8}}
           \put(-2,0){\line(-1,0){8}}
           \put(-10,-3){\fbox{Geometry 2}}
           \end{picture}
          }
\put(-10,-20){
           \begin{picture}(20,14)(-10,-3)
           \put(0,0){\circle*{0.5}}
           \put(2,0){\circle*{0.5}}
           \put(2,0){\line(1,0){8}}
           \put(0,2){\circle*{0.5}}
           \put(0,2){\line(0,1){8}}
           \put(0,-3){\fbox{Geometry 3}}
           \end{picture}
           }
\put(15,-30){
          \begin{picture}(20,24)(-10,-13)
          \put(0,0){\circle*{0.5}}
          \put(2,0){\circle*{0.5}}
          \put(2,0){\line(1,0){8}}
          \put(0,2){\circle*{0.5}}
          \put(0,2){\line(0,1){8}}
          \put(-2,0){\circle*{0.5}}
          \put(-2,0){\line(-1,0){8}}
          \put(0,-2){\circle*{0.5}}
          \put(0,-2){\line(0,-1){8}}
          \put(-10,-13){\fbox{Geometry 4}}
          \end{picture}
          }
\put(-10,-47){
           \begin{picture}(20,14)(-10,-3)
           \put(0,0){\circle*{0.5}}
           \put(2,0){\circle*{0.5}}
           \put(2,0){\line(1,0){8}}
           \put(0,2){\circle*{0.5}}
           \put(0,2){\line(0,1){8}}
           \put(1.2,0.4){\circle*{0.5}}
           \put(1.2,0.4){\line(3,1){8}}
           \put(0,-3){\fbox{Geometry 5}}
           \end{picture}
           }
\put(15,-57){
          \begin{picture}(20,24)(-10,-13)
          \put(0,0){\circle*{0.5}}
          \put(2,0){\circle*{0.5}}
          \put(2,0){\line(1,0){8}}
          \put(0,2){\circle*{0.5}}
          \put(0,2){\line(0,1){8}}
          \put(1.2,0.4){\circle*{0.5}}
          \put(1.2,0.4){\line(3,1){8}}
          \put(-1.2,-0.4){\circle*{0.5}}
          \put(-1.2,-0.4){\line(-3,-1){8}}
          \put(-2,0){\circle*{0.5}}
          \put(-2,0){\line(-1,0){8}}
          \put(0,-2){\circle*{0.5}}
          \put(0,-2){\line(0,-1){8}}
          \put(-10,-13){\fbox{Geometry 6}}
          \end{picture}
          }
\end{picture}
\caption{The six different geometries we have simulated. For each geometry
we have reported as a dot the starting point of the walk (located at the 
origin) and the starting points of the excluded needles (located at distance
two from the origin).}\label{picture}
\end{figure}

Notice that in the last case it is necessary that the needles start from a 
point located at distance two from the origin, 
otherwise the SAW starting
from the origin would necessarily touch one of the needles.

Let us now define a few observables which will allow us to study 
the effect of the introduction of the excluded region. If $(x,y,z)$ are 
the coordinates of the end-point of the SAW (or, equivalently,
$(r,\theta,\phi)$ in polar coordinates) a natural choice of 
observables is given by 
$r^l Y_{l,m}(\theta,\phi)$, $l\not=0$, where $Y_{l,m}(\theta,\phi)$ 
are the spherical harmonics. If $\< r^l Y_{l,m}(\theta,\phi)\>$
is not identically vanishing because of the residual cubic symmetry 
which survives the introduction of the excluded region, this quantity 
is a natural candidate for a direct determination of the exponent $\Delta$.
Indeed, since the critical limit is rotational-invariant,
the leading term for $N\to\infty$, i.e. $A(k)$ in formula \reff{coordinate},
will vanish and thus this quantity 
will scale as $N^{l\nu - \Delta}$. 

Let us now classify the various 
observables according to the value of $l$.
For $l=1$ the possibilities are 
\begin{eqnarray}
r(Y_{1,1}(\theta,\phi) + Y_{1,-1}(\theta,\phi)) & \sim & x \\
r(Y_{1,1}(\theta,\phi) - Y_{1,-1}(\theta,\phi)) & \sim & y  \\
r Y_{1,0} (\theta,\phi) & \sim  & z 
\end{eqnarray}
We can thus define $O_1(x) = x$ and the obvious permutations. This observable
is useful whenever the region $\cal R$ is not invariant under the inversion
of the $x$-axis. 

Let us now consider $l=2$. In this case we have
\begin{eqnarray}
r^2(Y_{2,2}(\theta,\phi) + Y_{2,-2}(\theta,\phi)) & \sim & x^2 - y^2\\
r^2(Y_{2,2}(\theta,\phi) - Y_{2,-2}(\theta,\phi)) & \sim & x y\\
r^2(Y_{2,1}(\theta,\phi) + Y_{2,-1}(\theta,\phi)) & \sim & x z\\
r^2(Y_{2,1}(\theta,\phi) - Y_{2,-1}(\theta,\phi)) & \sim & y z\\
r^2 Y_{2,0} (\theta,\phi) &\sim & x^2 + y^2 - 2 z^2
\end{eqnarray}
We can thus define three observables 
\begin{eqnarray}
O_{2,1} (x,y) &=& x^2 - y^2 \\
O_{2,2} (x,y) &=&  x y \\
O_{2,3} (x,y) &=& z^2-{1\over2} (x^2 + y^2) 
\end{eqnarray}
It is clear that, whenever a symmetry between two axes exists, 
the two variables $O_{2,1}$ and $O_{2,3}$ are equivalent. In our calculation
we have thus only considered the last two quantities.

We have finally considered $l=4$. In this case many different observables 
can be defined. We have only considered 
\begin{eqnarray}
O_{4,1} &=& {2\over3} \left[x^4 + y^4 + z^4 - 3 (x^2 y^2 + x^2 z^2 + y^2 z^2)
         \right] \\
O_{4,2}(x) &=& x^4 + \frac{1}{2}\left(y^4+z^4 \right)-3x^2\left(y^2+z^2\right)
\end{eqnarray}
Notice that when the excluded region is symmetric under any exchange of two
axes $O_{4,1}$ and $O_{4,2}$ are equivalent. It is also worth remarking
that $O_{4,1}$ and $O_{4,2}$ do not vanish 
even in absence of any excluded region, as they are invariant under 
the full cubic group.

Let us now define exactly our observables for the various geometries:
\begin{enumerate}
\item geometry 1 [needle along the positive $x$-axis]: we measure 
$O_1\equiv O_1(x)$, $O_2\equiv O_{2,3}(y,z)$ and $O_4\equiv O_{4,2}(x)$;
\item geometry 2 [two needles along the $x$-axis]: we measure  
$O_2\equiv O_{2,3}(y,z)$ and $O_4\equiv O_{4,2}(x)$;
\item geometry 3 [two needles along the positive $x$- and $y$-axis]:
we measure $O_1\equiv {1\over2} (O_1(x)+ O_1(y))$, 
$O_2\equiv {1\over2} (O_{2,3}(y,z)+O_{2,3}(z,x))=-{1\over2}O_{2,3}(x,y)$,
 $\tilde{O}_2\equiv O_{2,2}(x,y)$ and 
$O_4 \equiv O_{4,1}$;
\item geometry 4 [four needles in the $xy$-plane]: we measure 
$O_2\equiv {1\over2} (O_{2,3}(y,z)+O_{2,3}(z,x))=-{1\over2}
O_{2,3}(x,y)$ and $O_4 \equiv {1\over2} (O_{4,2}(x) +
O_{4,2}(y))$;
\item geometry 5 [three needles along three different axes]: 
we measure $O_1\equiv {1\over3} (O_1(x)+ O_1(y) + O_1(z))$, 
$\tilde{O}_2 \equiv \frac{1}{3}(O_{2,2}(x,y)+O_{2,2}(x,z)+O_{2,2}(x,y))$
and $O_4 \equiv O_{4,1}$;
\item geometry 6 [six needles]: we measure $O_4 \equiv O_{4,1}$.
\end{enumerate}

\section{The Monte Carlo simulation}

\subsection{Monte Carlo observables}

The purpose of our simulation was to compute the mean values of the 
observables we have defined in the previous Section in presence of an 
excluded region $\cal R$. A direct strategy would be to simulate 
SAWs in presence of the region $\cal R$ and then to compute the 
mean values of the various observables in the usual way. However 
this strategy requires different simulations for different excluded 
regions $\cal R$. To avoid repeating the runs many times we have simulated 
SAWs without any excluded region and then we have reweighted the 
results in order to obtain the mean values of interest. Since,
as we have previously discussed, the introduction of the excluded region
is an irrelevant perturbation, this strategy does not introduce any significant
loss of efficiency.

Let us suppose 
we want to compute $\<{\cal O}\>_{N,\cal R}$ where 
$\<\cdot \>_{N,\cal R}$ indicates the ensemble of SAWs of length $N$ 
that do not intersect the region $\cal R$. Then we simply use
\be
\<{\cal O}\>_{N,\cal R} \, =\, 
  { \<{\cal O}\, \theta^{({\cal R})}\>_N \over 
    \< \theta^{({\cal R})}\>_N }
\label{eq4.1}
\ee
where $\<\cdot \>_N$ indicates the average in the ensemble of all 
SAWs of length $N$
and $\theta^{\cal R}$ is an observable which assumes the value one if a 
walk does not intersect $\cal R$ and zero otherwise. Notice moreover 
that $\< \theta^{({\cal R})}\>_N$ gives also the probability $p_{\cal R}$
that a SAW intersects the excluded region as 
$p_{\cal R} = 1 - \< \theta^{({\cal R})}\>_N$.

We have used \reff{eq4.1} to obtain from our simulation the mean values 
$\<{\cal O}\>_{N,\cal R}$: indeed it is enough, beside measuring at each 
Monte Carlo step $i$ the value ${\cal O}_i$, to record 
$\theta^{({\cal R})}_i$ which assumes the value one if the walk 
intersects $\cal R$, zero otherwise and then estimate 
$\<{\cal O}\>_{N,\cal R}$ by
\be
{\sum_{i=1}^n {\cal O}_i\, \theta^{({\cal R})}_i \over 
 \sum_{i=1}^n  \theta^{({\cal R})}_i} \;\; .
\ee
In order to further reduce the variance of our estimates
we have also symmetrized our observables. To clarify the method 
suppose we want to compute $\<O_1\>_{N,{\cal R}}$ for geometry 1
(${\cal R}$ is the positive $x$-axis).  We could consider 
\be 
  {\< x \theta^{(+x)}\>_N \over \< \theta^{(+x)}\>_N} \;\; .
\ee
A symmetrized alternative is 
\be
{\< x (\theta^{(+x)} - \theta^{(-x)}) + 
    y (\theta^{(+y)} - \theta^{(-y)}) + 
    z (\theta^{(+z)} - \theta^{(-z)}) \>_N \over 
 \< \theta^{(+x)} + \theta^{(-x)} + 
    \theta^{(+y)} + \theta^{(-y)} +
    \theta^{(+z)} + \theta^{(-z)} \>_N } \;\; .
\ee
It is obvious that this quantity has the same mean value of the previous 
one. However this symmetrization reduces the error on the estimates
essentially at no computational cost. 
Indeed, since we study all six geometries at the same time, 
we must check in any case if the walk intersects any of the axes.

In general, given a geometry $\cal R$ we have considered all the possible
sets $s_{\cal R}$ of excluded needles which can be obtained by all the
transformations of the cubic group
(six for geometry 1, three for geometry 2, twelve for
geometry 3, three for geometry 4, eight for geometry 5, one for geometry 6).
To each region $s_{\cal R}$ we associate a variable
${{\theta}}^{\left(s_{\cal R}\right)}$, 
which is one if the walk does not intersect the set $s_{\cal R}$, zero
otherwise and ${O}^{\left(s_{\cal R}\right)}$ which is the suitably 
transformed variable. Then we compute $\< O\>_{N,\cal R}$ from
\be
\langle O \rangle_{N,\cal R} = 
\frac{\langle\displaystyle\sum_{\{s_{\cal R}\}}
     \theta^{\left(s_{\cal R}\right)}
     O^{\left(s_{\cal R}\right)}\rangle}
    {\langle \displaystyle\sum_{\{s_{\cal R}\}}
   \theta^{\left(s_{\cal R}\right)}\rangle} 
\ee
The symmetrization has a large effect on the static variances of the various 
observables. For instance consider for each geometry the ratio 
\be
R\, =\, {\hbox{var}\ \theta^{\cal R} \over 
         \hbox{var}\ {1\over n_{\cal R} }
            \sum {\theta}^{\left(s_{\cal R}\right)} }
\ee
where ``var" indicates the static variance and $n_{\cal R}$ is the 
number of terms in the sum. We find 
$R = 6.6$, 3.2, 3.8, 2.7, 1.8, 1 in the six geometries, which 
represents a considerable improvement. The improvement on the error
bars is however not so large as the symmetrized observables are 
more correlated than the unsymmetrized ones. Thus the error bars are only 
20-30\% better.

To conclude let us comment briefly on the determination of the error bars.
{}From a Monte Carlo run of $n$ iterations we have estimated 
$\langle O \rangle_{N,\cal R}$ using
\be 
\langle O \rangle_{N,\cal R} \, =\, 
\frac{\overline{\displaystyle\sum_{\{s_{\cal R}\}}
     \theta^{\left(s_{\cal R}\right)}
     O^{\left(s_{\cal R}\right)}}}
    {\overline{\displaystyle\sum_{\{s_{\cal R}\}}
   \theta^{\left(s_{\cal R}\right)}} }
\ee
where, as usual $\overline{\cal O} = {1\over n}\sum_i {\cal O}_i$. To compute 
the error bars one must take into account the correlation between
the numerator and the denominator: the independent error formula 
indeed overestimates the true error bar. For $O_1$ the difference is of about 
20$-$30\%, for $O_2$ and $\tilde{O}_2$ of 4$-$7\%, while for $O_4$ the 
difference is negligible.
We have thus used the following relation, valid in the  large 
sample limit,
\be
\hbox{Var}\left(\frac{\overline{\displaystyle\sum_{\{s_{\cal R}\}}
\theta^{\left(s_{\cal R}\right)}O^{\left(s_{\cal R}\right)}}}
{\overline{\displaystyle\sum_{\{s_{\cal R}\}}
{\theta}^{\left(s_{\cal R}\right)}}}\right)
      =    \frac{\langle\displaystyle\sum_{\{s_{\cal R}\}}
\theta^{\left(s_{\cal R}\right)}O^{\left(s_{\cal R}\right)}
\rangle^2}{\langle \displaystyle\sum_{\{s_{\cal R}\}}
\theta^{\left(s_{\cal R}\right)}\rangle^2}
\hbox{Var}\left({\cal A}\right).
\ee
where $\cal A$ is given by
\be
{\cal A} =
     \frac{\displaystyle\sum_{\{s_{\cal R}\}}
     \theta^{\left(s_{\cal R}\right)}O^{\left(s_{\cal R}\right)}}
   {\langle \displaystyle\sum_{\{s_{\cal R}\}}\theta^{\left(s_{\cal R}\right)}
     O^{\left(s_{\cal R}\right)}\rangle} -
  \frac{\displaystyle\sum_{\{s_{\cal R}\}}{\theta}^{\left(s_{\cal R}\right)}}
     {\langle \displaystyle\sum_{\{s_{\cal R}\}}
     {\theta}^{\left(s_{\cal R}\right)}\rangle}
\label{calA}
\ee
The variance of $\cal A$ has been computed using the standard techniques of 
autocorrelation analysis \cite{Sokal1,Sokal2}. 
Since the autocorrelation function has a very long tail
(due to the fact that $\tau_{exp}\sim N/f \gg \tau_{int,X}$ 
where $f$ is the acceptance fraction of the algorithm and $X$ 
a generic {\em global} observable), the self-consistent windowing
method proposed in \cite{Sokal1} does not work. Indeed even using 
a large window of $40 \tau_{int,{\cal A}}$, the autocorrelation time 
is largely underestimated. To get a reliable estimate of 
$\tau_{int}$ we have used the recipe proposed in the Appendix of 
\cite{Sokal2}. We compute $\tau_{int,X}$ by
\begin{eqnarray}
{\tau'}_{int,X}(M) &=& {1\over 2} + \sum_{t=1}^M \rho_X(t) \\
\tau_{int,X} &=& {\tau'}_{int,X}(M) + \rho_X(M) M \log
    \left( {N\over Mf}\right)
\label{taunew}
\end{eqnarray}
where $\rho_X(t)$ is the normalized autocorrelation function,
$N$ the length of the walk, $f$ the acceptance fraction; 
$M$ is determined self-consistently and is the smallest integer such that 
$20 {\tau'}_{int,X}(M) > M$. We have checked the {\em ad hoc} definition 
\reff{taunew} in the case of ordinary random walks for which 
exact results are available \cite{Sokal1}: the error in the estimation 
of the tail turns out to be at most 10\%. We have thus set the error
bars on the autocorrelation times 
adding to the error in the determination of ${\tau'}_{int,X}(M)$, one tenth 
of the last term in \reff{taunew}.

\subsection{The algorithm}

We have simulated SAWs of fixed length $N$ in three dimensions without 
any excluded region $\cal R$. The walk is given by $N+1$ lattice 
points $\{\omega_{i}\}$, $i=0,\ldots,N$, and always starts from the 
origin, i.e. $\omega_0=0$. Since all our observables are completely
symmetric under the cubic group, it is not restrictive to fix the 
first step; we have chosen $\omega_1=(1,0,0)$.

The simulation used the {\em pivot} algorithm~\cite{Lal,MacDonald,Sokal1}. 
In the standard implementation
a site $\omega_k$ in the walk  and an element
$g$ of the lattice symmetry 
group are chosen randomly, with uniform probability.
The proposed configuration, obtained from the actual one by applying $g$ to 
the part of the walk subsequent to $\omega_k$, is accepted
whenever self-avoiding. 
This algorithm is extraordinarily efficient for the study of global 
observables like, for example, the end-point position; 
indeed, the integrated autocorrelation time for these 
observables grows with the number 
$N$ of steps in the walk like $N^p$, where $p\approx 0.11$ in $d=3$,
while the CPU-time to produce an independent walk scales like $N$ 
which is the optimal situation.
\begin{table}
\begin{center}
\begin{tabular}{|c|c||c|c|}
\hline
& & standard pivot & improved pivot  \\
\hline \hline
geometry &  observable & $\tau_{int}$ & $\tau_{int}$  \\
\hline \hline
1 &  $O_{1}$ & $6.92 \pm 0.23 $ & $ 1.4869 \pm  $ 0.0030\\
\cline{2-4}
& $p_{\cal R}$ & $ 282 \pm 60 $ &  $ 5.381 \pm 0.013$ \\
\hline
2 & $p_{\cal R}$ & $280 \pm 60 $ & $  5.026\pm 0.011 $ \\
\hline
3 & $O_{1}$ & $ 7.73\pm 0.27 $ & $ 1.4946 \pm 0.0030 $ \\
\cline{2-4}
& $p_{\cal R}$ & $ 307 \pm 68 $ & $ 5.081 \pm 0.012 $\\
\hline
4 & $p_{\cal R}$ & $ 311 \pm 69 $ & $ 4.2485\pm  0.0083$ \\
\hline
5 & $O_{1}$ & $ 8.58 \pm 0.32 $ & $  1.4293\pm 0.0029 $\\
\cline{2-4}
& $p_{\cal R}$ & $ 312  \pm 70 $ & $  4.628 \pm 0.010 $ \\
\hline 
6 &  $p_{\cal R}$ & $ 261 \pm 53 $ & $ 2.7910 \pm 0.0048 $\\
\hline 
\end{tabular}
\end{center}
\caption{Autocorrelation times for the pivot algorithm and 
our improvement for the observables $O_1$ (in the geometries 
where it does not vanish) and $p_{\cal R}$.
Here $N=1000$.}
\label{auto}
\end{table}

However in our case not all the observables are of global character: 
an example is the probability of intersection $p_{\cal R}$. Indeed, as 
can be seen from table \ref{auto}, already at $N=1000$, 
$\tau_{int,p_{\cal R}}\approx 300$. It is easy to understand the origin of 
these autocorrelation times: indeed suppose the walk intersects ${\cal R}$
and let $\alpha$ be the smallest integer such that $\omega_\alpha\in \cal R$.
Then all subsequent walks in the simulation will also intersect ${\cal R}$
at least until a successful pivot move at point 
$i$ with $i<\alpha$ is performed. The probability of such a move 
is $\alpha f/N$ where $f$ is the acceptance fraction. Now the problem 
is that the typical $\alpha$ is very small: if one considers the set of walks
that intersect at least one of the six needles of the geometry 6, the 
mean value of the smallest $\alpha$ such that $\omega_\alpha\in \cal R$
is $\approx 13$, with very small $N$-dependent corrections. 
Thus a rough estimate of the autocorrelation times 
should be $N/(13 f)$. For $N=1000$ we have $f\approx 0.45$ so that we expect
$\tau\approx 200$ which is indeed the order of magnitude we find. 
Moreover we expect $\tau$ to increase as $N^{\approx 1}$ as 
it should for local observables. 
Indeed we find that $\tau$ for $p_{\cal R}$ increases as $N^p$ with 
$p \approx 0.9$ except in the case of geometry 6 where we observe 
$p\approx 0.6$, which could however well be an effective 
exponent in the region $500\le N\le 32000$. 

The other observables, like $O_1$, are instead of more global character
and indeed the autocorrelation times are much smaller\footnote{When
we speak of autocorrelation time for $O_1$ we really intend the 
$\tau_{int,{\cal A}}$ as defined in \protect\reff{calA} as this is 
the observable which controls the error on $O_1$.}. They also increase 
more slowly with $N$; for $O_1$ we find $\tau_{int,O_1}\sim N^{\approx 0.3}$.

To eliminate these long autocorrelation times one must increase the 
frequency of the moves with the first points of the walk as pivots.
We have thus modified the algorithm in the following way: 
one iteration consists now of the following three moves:
\begin{itemize}
\item one move with pivot point in $\omega_1$:
\item one move with pivot point $\omega_k$ with $k$ uniformly chosen 
      in the interval $1<k\le 13$;
\item one move with pivot point $\omega_k$ with $k$ uniformly chosen 
      in the interval $13<k\le N-1$.
\end{itemize}
With this improvement, the autocorrelation times, are greatly reduced. 
Table~\ref{auto} contains a comparison between autocorrelation time for 
the two algorithms for walks of length $ N_{tot} = 1000 $.
We observed sensible reductions of autocorrelation times
for each observable, but the most impressive results are obtained for the
observable $p_{\cal R}$, which denotes the fraction of
walks that intersect the considered excluded set.
To compare CPU-times one should notice 
that one iteration of the improved algorithm takes three times the CPU-time 
of an iteration of the standard algorithm. Thus in practice use of the 
improved algorithm allowed us to gain a factor of $15-20$ on $p_{\cal R}$
(although no improvement on the exponent $p$) and a factor 
$1.5-2$ on global observables like $O_1$.
\begin{table}
\begin{center}
\begin{small}
\begin{tabular}{|c||c|c|c|c|c|c|c|}
\hline
& 500 &1000 & 2000 & 4000 & 8000 & 16000 & 32000 \\
\hline\hline
$N_{iter.}$ &$ 3\cdot10^7$ & $5\cdot10^7$ &$ 6\cdot10^7$
& $7\cdot10^7$ & $1\cdot10^8$ & $5\cdot10^7$&
$2\cdot10^7 $\\
\hline
\end{tabular}
\end{small}
\end{center}
\caption{Number of iterations. }
\label{iterazioni}
\end{table}

\section{The results}

We have studied SAWs with length $500\le N \le 32000$. The number of iterations
for each value of $N$ is reported in table \ref{iterazioni}.
The total simulation required 2500 hours of CPU on a AlphaStation 600 
Mod 5/266. We have computed $\Delta$ 
studying the quantities we have previously discussed, whose leading behaviour
for $N\to \infty$ is
\be
\< O_k \>_N =  B_{\cal R}(k) \,N^{\sigma}
\label{Oleading}
\ee
where $\sigma = k \nu - \Delta$, 
$k=1,2,4$. We have performed standard power-law fits neglecting the 
next subleading
terms. This introduces a systematic error which in our case could be 
particularly serious due to small value of $\Delta$. Indeed one 
expects corrections to \reff{Oleading} of the form 
$N^{k\nu -2\Delta}$ which decay very slowly and could thus give 
sizable corrections even at the relatively large values of $N$ we use.
To get an idea of the systematic error we have repeated the fits 
using each time only those values of $N$ satisfying $N\ge N_{cut}$. In this way
one gets different estimates which should converge to $\Delta$ 
as $N_{cut}$ goes to infinity. If the different estimates 
are essentially independent of $N_{cut}$ (within error bars)
one can reasonably trust the estimate of $\Delta$, otherwise one gets an
idea of the size of the systematic error.

We show the results of our fits for each geometry in 
Tables~\ref{prima}--\ref{ultima}.
In the first column one can find 
the different values of $N_{cut}$, in the second one the 
raw Monte Carlo data, in the next one
the estimated value of the exponent and finally the
confidence level of the fit. In the last column we report
the ``classical prediction" for the 
exponent $\sigma$, obtained using \reff{Deltapredicted}.
\begin{table}
\begin{center}
\begin{small}
\begin{tabular}{|c||c|c|c|r|l|}
\hline
\multicolumn{6}{|c|}{geometry 1}\\
\hline\hline
 & cut &$ \<O\> \pm \delta\<O\>$  & $\sigma \pm \delta\sigma $ & $ CL$ & \\
\hline\hline
 &   500 & $              -1.5259 \pm               0 .0014 $ &
 $   0.40044 \pm   0.00044 $ &  $ 9\cdot10^{-7}$\% &\\
\cline{2-5}
 &  1000 & $              -2.0229 \pm                0.0018 $ &
 $   0.39831 \pm   0.00060 $ &   0.114\% &\\
\cline{2-5}
 &  2000 & $              -2.6751 \pm                0.0028 $ &
 $   0.39584 \pm   0.00091 $ & 14.118\% & $\nu -\Delta =$\\
\cline{2-5}
 $O_1$&  4000 & $              -3.5322 \pm               0 .0044 $ &
 $   0.3932 \pm   0.0016 $ & 62.627\% &$1-\nu =$\\
\cline{2-5}
 &  8000 & $              -4.6324 \pm                0.0060 $ &
 $   0.3955 \pm   0.0029 $ & 92.618\% & $ 0.4123 \pm 0.0006$\\
\cline{2-5}
 & 16000 & $              -6.092 \pm 0               .015 $ &
 $   0.3962 \pm   0.0080 $ & &   \\
\cline{2-5}
 & 32000 & $              -8.018 \pm  0              .040 $  &&& \\

\hline\hline
 &   500 & $             -20.457 \pm   0             .047 $ &
 $   0.97813 \pm   0.00098 $ &   $29\cdot10^{-4}$\%&\\
\cline{2-5}
 &  1000 & $             -39.944 \pm 0               .089 $ &
 $   0.9818 \pm   0.0013 $ &  3.622\%&\\
\cline{2-5}
 &  2000 & $             -78.25 \pm  0              .20 $ &
 $   0.9859 \pm   0.0020 $ & 53.836\%& $ 2\nu -\Delta= 1 $\\
\cline{2-5}
$O_2$ &  4000 & $            -155.00 \pm  0              .43 $ &
 $   0.9866 \pm   0.0031 $ & 35.367\%& \\
\cline{2-5}
 &  8000 & $            -306.42 \pm                0.85 $ &
 $   0.9908 \pm   0.0054 $ & 26.964\%&\\
\cline{2-5}
 & 16000 & $            -606.0 \pm               3.0 $ &
 $  1.005 \pm   0.014 $ &  & \\
\cline{2-5}
 & 32000 & $           -1217. \pm              10. $ &&&\\
\hline\hline
 &   500 & $         ( -110 \pm            11)\cdot10^2 $ &
 $  2.277 \pm   0.035 $ & 15.228\%&\\
\cline{2-5}
 &  1000 & $          (-596 \pm            50)\cdot10^2 $ &
 $  2.239 \pm   0.044 $ & 20.112\%&\\
\cline{2-5}
 &  2000 & $         (-295 \pm           25)\cdot10^3 $ &
 $  2.182 \pm   0.063 $ & 22.338\%& $4\nu - \Delta=$\\
\cline{2-5}
 $O_{4}$ &  4000 & $       ( -160 \pm          12)\cdot10^4 $ &
 $  2.056 \pm   0.096 $ & 51.065\%& $1 + 2\nu =$\\
\cline{2-5}
 &  8000 & $      (  -642 \pm          53)\cdot10^4 $ &
 $  2.10 \pm   0.20 $ & 25.931\%& $2.1754\pm 0.0012$ \\
\cline{2-5}
 & 16000 & $       (-304 \pm         41)\cdot10^5 $ &
 $  1.43 \pm   0.62 $ & & \\
\cline{2-5}
 & 32000 & $      ( -82 \pm        34)\cdot10^6 $ &&& \\
\cline{2-5}
\hline
\end{tabular}
\end{small}
\end{center}
\caption{One excluded needle. Here $CL$ denotes the confidence level of the 
fit.} \label{prima}
\end{table}
\begin{table}
\begin{center}
\begin{small}
\begin{tabular}{|c||c|c|c|r|l|}
\hline
\multicolumn{6}{|c|}{geometry 2}\\
\hline\hline
 & cut &$ \<O\> \pm \delta\<O\>$  & $\sigma \pm \delta\sigma $ & $ CL$ & \\
\hline\hline
 &   500 & $             -45.97 \pm                0.11 $ &
 $  0 .9622 \pm  0 .0010 $ & $22\cdot 10^{-5}$ \% &\\
\cline{2-5}
 &  1000 & $             -88.71 \pm                0.20 $ &
 $   0.9664 \pm  0 .0014 $ &  1.206\% &\\
\cline{2-5}
 &  2000 & $            -171.86 \pm                0.46 $ &
 $   0.9712 \pm  0 .0021 $ & 32.894\% & $ 2\nu - \Delta= 1$\\
\cline{2-5}
 $O_{2}$ &  4000 & $            -336.7 \pm               1.0 $ &
 $   0.9732 \pm  0 .0034 $ & 23.052\% & \\
\cline{2-5}
 &  8000 & $            -658.2 \pm               2.0 $ &
 $   0.9801 \pm  0 .0060 $ & 33.624\% &\\
\cline{2-5}
 & 16000 & $           -1292.7 \pm               6.9 $ &
 $   0.994 \pm  0 .016 $ &   &\\
\cline{2-5}
 & 32000 & $           -2575. \pm              25. $ &&&\\
\hline\hline
 &   500 & $          (-299 \pm            12)\cdot 10^2 $ &
 $  2.206 \pm  0.015 $ & 11.869\%   & \\
\cline{2-5}
 &  1000 & $         (-1437 \pm            50)\cdot 10^2 $ &
 $  2.190 \pm   0.019 $ & 14.368\% &\\
\cline{2-5}
 &  2000 & $         (-670 \pm           25)\cdot 10^3$ &
 $  2.162 \pm   0.028 $ & 15.761\% & $ 4\nu - \Delta= $ \\
\cline{2-5}
 $O_{4}$  &  4000 & $        (-328 \pm          12)\cdot 10^4 $ &
 $  2.099 \pm   0.045 $ & 39.275\% &$1 + 2\nu=$  \\
\cline{2-5}
 &  8000 & $       (-1374 \pm          54)\cdot 10^4 $ &
 $  2.122 \pm   0.085 $ & 18.371\% &$2.1754 \pm 0.0012 $\\
\cline{2-5}
 & 16000 & $       (-634 \pm         40)\cdot 10^5 $ &
 $  1.80 \pm   0.26 $ &   &\\
\cline{2-5}
 & 32000 & $      (-221 \pm        37)\cdot 10^6 $ &&& \\
\hline
\end{tabular}
\end{small}
\end{center}
\caption{Two opposite needles. Here $CL$ denotes the confidence level of the
fit.}
\end{table}
\begin{table}
\begin{center}
\begin{small}
\begin{tabular}{|c||c|c|c|r|l|}
\hline
\multicolumn{6}{|c|}{geometry 3}\\
\hline\hline
 & cut &$ \<O\> \pm \delta\<O\>$  & $\sigma \pm \delta\sigma $ & $CL$ & \\
\hline\hline
 &   500 & $              -1.6573 \pm                0.0015 $ &
 $   0.39267 \pm   0.00044 $ & $4\cdot 10^{-8}$  \%&\\
\cline{2-5}
 &  1000 & $              -2.1859 \pm               0 .0020 $ &
 $   0.39038 \pm   0.00060 $ &   0.037\% &\\
\cline{2-5}
 &  2000 & $              -2.8761 \pm                0.0030 $ &
 $   0.38767 \pm   0.00092 $ & 13.770\% &$\nu - \Delta$\\
\cline{2-5}
 $O_{1}$ &  4000 & $              -3.7761 \pm                0.0048 $ &
 $   0.3850 \pm  0 .0016 $ & 59.748\% & $1 - \nu = $\\
\cline{2-5}
 &  8000 & $              -4.9246 \pm               0 .0065 $ &
 $   0.3873 \pm  0 .0029 $ & 70.556\%  &$
0.4123 \pm 0.0006 $\\
\cline{2-5}
 & 16000 & $              -6.436 \pm                0.016 $ &
 $   0.3901 \pm   0.0081 $ &   &   \\
\cline{2-5}
 & 32000 & $              -8.435 \pm                0.042 $ &&&\\

\hline\hline
 &   500 & $             -10.696 \pm               0 .025 $ &
 $   0.9717 \pm  0 .0010 $ &   $5\cdot 10^{-5}$\% & \\
\cline{2-5}
 &  1000 & $             -20.740 \pm               0 .047 $ &
 $   0.9761 \pm  0 .0013 $ &  2.063\%&\\
\cline{2-5}
 &  2000 & $             -40.42 \pm                0.11 $ &
 $  0 .9809 \pm   0.0020 $ & 55.682\%& $2\nu -\Delta = 1
$\\
\cline{2-5}
 $O_2 $&  4000 & $             -79.86 \pm                0.24 $ &
 $   0.9811 \pm   0.0034 $ & 35.520\% & \\
\cline{2-5}
 &  8000 & $            -157.19 \pm               0 .45 $ &
 $   0.9859 \pm   0.0057 $ & 31.462\% &\\
\cline{2-5}
 & 16000 & $            -309.9 \pm               1.6 $ &
 $  1.000 \pm   0.015 $ &   &\\
\cline{2-5}
 & 32000 & $            -619.8 \pm               5.6 $ &&&\\

\hline\hline
 &   500 & $               -0.1650 \pm                0.0081 $ &
 $   0.772 \pm   0.036 $ &  5.684\% &\\
\cline{2-5}
 &  1000 & $               -0.276 \pm                0.016 $ &
 $   0.730 \pm   0.057 $ &  4.327\% &\\
\cline{2-5}
 &  2000 & $               -0.572 \pm                0.035 $ &
 $   0.512 \pm   0.092 $ & 89.526\% & $2\nu -\Delta =1$\\
\cline{2-5}
 $\tilde{O}_2$ &  4000 & $              -0.819 \pm                0.079 $ &
 $   0.53 \pm   0.18 $ & 74.224\% &\\
\cline{2-5}
 &  8000 & $              -1.07 \pm                0.18 $ &
 $   0.77 \pm   0.43 $ & 63.693\% &\\
\cline{2-5}
 & 16000 & $              -2.02 \pm                0.67 $ &
 $   0.1 \pm  1.4 $ &   &\\
\cline{2-5}
 & 32000 & $              -2.2 \pm               2.0 $ &&&\\

\hline\hline
 &   500 & $          (-166 \pm            12)\cdot10^2 $ &
 $  2.244 \pm   0.025 $ & 10.182\%&\\
\cline{2-5}
 &  1000 & $          (-864 \pm            52)\cdot10^2 $ &
 $  2.211 \pm   0.032 $ & 16.821\%&\\
\cline{2-5}
 &  2000 & $         (-414 \pm           25)\cdot10^3 $ &
 $  2.171 \pm   0.046 $ & 17.299\%&$4\nu -\Delta
=$\\
\cline{2-5}
 $O_4$&  4000 & $        (-214 \pm          12)\cdot10^4 $ &
 $  2.071 \pm   0.072 $ & 40.820\%&$1 + 2 \nu = $\\
\cline{2-5}
 &  8000 & $        (-871 \pm          54)\cdot10^4 $ &
 $  2.10 \pm   0.14 $ & 18.912\%& $ 2.1754 \pm 0.0012 $\\
\cline{2-5}
 & 16000 & $       (-409 \pm         42)\cdot10^5 $ &
 $  1.56 \pm   0.44 $ &   &\\
\cline{2-5}
 & 32000 & $      (-120 \pm        34)\cdot10^6 $ &&&\\
\hline
\end{tabular}
\end{small}
\end{center}
\caption{Two needles along two different axes. Here $CL$ denotes the 
confidence level of the fit.}
\end{table}
\begin{table}
\begin{center}
\begin{small}
\begin{tabular}{|c||c|c|c|r|l|}
\hline
\multicolumn{6}{|c|}{geometry 4}\\
\hline\hline
 & cut &$ \<O\> \pm \delta\<O\>$  & $\sigma \pm \delta\sigma $ & $ CL$ & \\
\hline\hline
 &   500 & $             -25.876 \pm                0.071 $ &
 $   0.9439 \pm  0 .0012 $ & $  10^{-7}$\% &\\
\cline{2-5}
 &  1000 & $             -49.00 \pm                0.13 $ &
 $   0.9501 \pm   0.0016 $ &   0.747\% &\\
\cline{2-5}
 &  2000 & $             -93.52 \pm                0.29 $ &
 $   0.9561 \pm   0.0024 $ & 34.436\% &$ 2\nu -
\Delta = 1 $ \\
\cline{2-5}
 $O_{2}$ &  4000 & $            -182.09 \pm                0.63 $ &
 $   0.9552 \pm   0.0040 $ & 19.884\% &\\
\cline{2-5}
 &  8000 & $            -351.1 \pm               1.2 $ &
 $   0.9643 \pm   0.0072 $ & 35.181\% &\\
\cline{2-5}
 & 16000 & $            -682.1 \pm               4.1 $ &
 $   0.981 \pm   0.020 $ &   &\\
\cline{2-5}
 & 32000 & $           -1347. \pm              17. $ &&&\\

\hline\hline
 &   500 & $          (-518 \pm            13)\cdot10^2 $ &
 $  2.1760 \pm  0 .0099 $ &  3.475\%&\\
\cline{2-5}
 &  1000 & $         (-2425 \pm            55)\cdot10^2 $ &
 $  2.161 \pm  0 .013 $ &  6.562\%&\\
\cline{2-5}
 &  2000 & $        (-1104 \pm           27)\cdot10^3 $ &
 $  2.142 \pm  0 .019 $ &  7.781\%&$ 4\nu - \Delta $\\
\cline{2-5}
 $O_{4}$ &  4000 & $        (-521 \pm          14)\cdot10^4 $ &
 $  2.094 \pm  0 .031 $ & 24.042\%&$1 +  2 \nu = $\\
\cline{2-5}
 &  8000 & $      ( -2181 \pm          61)\cdot10^4$ &
 $  2.1083 \pm  0 .057 $ &  9.664\%&$ 2.1754 \pm 0.0012 $ \\
\cline{2-5}
 & 16000 & $      ( -991\pm         42)\cdot10^5 $ &
 $  1.86 \pm   0.16 $ &&   \\
\cline{2-5}
 & 32000 & $      (-359 \pm        37)\cdot10^6 $ &&&\\
\hline
\end{tabular}
\end{small}
\end{center}
\caption{Four needles in a plane. Here $CL$ denotes the 
confidence level of the fit.}
\end{table}
\begin{table}
\begin{center}
\begin{small}
\begin{tabular}{|c||c|c|c|r|l|}
\hline
\multicolumn{6}{|c|}{geometry 5}\\
\hline\hline
 & cut &$ \<O\> \pm \delta\<O\>$  & $\sigma \pm \delta\sigma $ & $ CL$ & \\
\hline\hline
&   500 & $              -1.8113 \pm                0.0017 $ &
 $   0.38443 \pm   0.00045 $ &   $13\cdot10^{-10}$\% &\\
\cline{2-5}
 &  1000 & $              -2.3764 \pm                0.0022 $ &
 $   0.38193 \pm   0.00061 $ &   0.012\% &\\
\cline{2-5}
 &  2000 & $              -3.1097 \pm                0.0034 $ &
 $  0.37900 \pm   0.00093 $ & 12.449\%&$ \nu - \Delta=
$\\
\cline{2-5}
 $O_{1}$&  4000 & $              -4.0587 \pm                0.0053 $ &
 $   0.3762 \pm   0.0016 $ & 55.622\%&$1 - \nu = $\\
\cline{2-5}
 &  8000 & $              -5.2613 \pm                0.0073 $ &
 $   0.3786 \pm   0.0030 $ & 58.046\%& $ 0.4123 \pm
0.0006$\\
\cline{2-5}
 & 16000 & $              -6.833 \pm               0 .016 $ &
 $   0.3828 \pm   0.0082 $ &   &\\
\cline{2-5}
 & 32000 & $              -8.909 \pm                0.046 $ &&&\\

\hline\hline
&   500 & $               -0.1892 \pm                0.0088 $ &
 $   0.763 \pm   0.034 $ &  3.242\% &\\
\cline{2-5}
 &  1000 & $               -0.311 \pm               0 .017 $ &
 $   0.732 \pm   0.052 $ &  2.100\%&\\
\cline{2-5}
 &  2000 & $               -0.645 \pm                0.038 $ &
 $   0.518 \pm   0.084 $ & 81.853\%& $ 2\nu - \Delta= 1 $\\
\cline{2-5}
 $\tilde{O}_2$ &  4000 & $               -0.952 \pm                0.086 $ &
 $   0.50 \pm   0.17 $ & 63.392\% & \\
\cline{2-5}
 &  8000 & $              -1.20 \pm                0.17 $ &
$   0.79 \pm   0.39 $ & 61.993\%& \\
\cline{2-5}
 & 16000 & $              -2.27 \pm                0.66 $ &
 $   0.2 \pm  1.3 $ &   &\\
\cline{2-5}
 & 32000 & $              -2.5 \pm               2.2 $ &&&\\

\hline\hline
&   500 & $         ( -282 \pm            12)\cdot 10^2 $ &
 $  2.211 \pm   0.016 $ &  5.575\%&\\
\cline{2-5}
 &  1000 & $         (-1393 \pm            53)\cdot 10^2  $ &
 $  2.188 \pm   0.021 $ & 10.689\%&\\
\cline{2-5}
 &  2000 & $         (-649 \pm           26)\cdot 10^3 $ &
 $  2.159 \pm   0.030 $ & 11.283\%& $ 4\nu - \Delta= $\\
\cline{2-5}
 $O_4$&  4000 & $        (-321 \pm          13)\cdot 10^4 $ &
 $  2.088 \pm   0.048 $ & 30.950\%&$ 1 + 2 \nu =$ \\
\cline{2-5}
 &  8000 & $       (-1334 \pm          54)\cdot 10^4 $ &
 $  2.103 \pm   0.089 $ & 12.901\%& $2.1754 \pm 0.0012 $ \\
\cline{2-5}
 & 16000 & $       (-618 \pm         42)\cdot 10^5 $ &
 $  1.73 \pm   0.26 $ &   &\\
\cline{2-5}
 & 32000 & $      (-205 \pm        35)\cdot 10^6 $ &&&\\
\hline
\end{tabular}
\end{small}
\end{center}
\caption{Three needles along three different axes. Here $CL$ denotes 
the confidence level of the fit.}
\end{table}
\begin{table}
\begin{center}
\begin{small}
\begin{tabular}{|c||c|c|c|r|l|}
\hline
\multicolumn{6}{|c|}{geometry 6}\\
\hline\hline
 & cut &$ \<O\> \pm \delta\<O\>$  & $\sigma \pm \delta\sigma $ & $ CL$ & \\
\hline\hline
 &   500 & $          (-787 \pm            16)\cdot 10^2 $ &
 $  2.1540 \pm  0 .0086 $ &  4.443\%&\\
\cline{2-5}
 &  1000 & $         (-3606 \pm            67)\cdot 10^2  $ &
 $  2.141 \pm   0.011 $ &  7.879\%& \\
\cline{2-5}
 &  2000 & $        (-1611 \pm           35)\cdot 10^3 $ &
 $  2.123 \pm   0.017 $ &  9.185\%& $ 4\nu -
\Delta= $ \\
\cline{2-5}
 $O_4$&  4000 & $        (-746 \pm          17)\cdot 10^4 $ &
 $  2.078 \pm   0.028 $ & 33.022\%& $ 1 + 2 \nu = $ \\
\cline{2-5}
 &  8000 & $       (-3093 \pm          76)\cdot 10^4 $ &
 $  2.097 \pm   0.053 $ & 15.301\%& $2.1754 \pm 0.0012 $\\
\cline{2-5}
 & 16000 & $      (-1382\pm         56)\cdot 10^5 $ &
 $  1.89 \pm   0.15 $ &   &\\
\cline{2-5}
 & 32000 & $      (-513 \pm        50)\cdot 10^6 $ &&&\\
\hline
\end{tabular}
\end{small}
\end{center}
\caption{Six needles along the two directions of all the three 
axes. Here $CL$ denotes the 
confidence level of the fit.}\label{ultima}
\end{table}
In the computation of the expectation values we used the estimate for 
$\nu$ given by~\cite{Sokal2}
\be
\nu=0.5877 \pm 0.0006
\ee
Let us now comment on the results. Let us consider first the observable
$O_1$. Since the estimates for $N_{cut}\ge 4000$ are essentially
constant within error bars one could estimate
\be
\sigma = \cases{0.396(3) & for geometry 1 \cr
                0.388(3) & for geometry 3 \cr
                0.379(3) & for geometry 5 \cr}
\ee
These estimates have very small statistical error bars. However the fact
that the estimates do not agree within the stated errors is a clear
indication that there is a much larger error due to the 
neglected corrections to scaling. Indeed a closer look to the data 
shows that in all cases there is still an upward trend, although not
statistically significant as the change of $\sigma$ is smaller
than the error bars. It is thus more cautious to interpret the results 
for $\sigma$ as lower bounds on the true value. Thus we estimate
$\sigma = 0.39\gtapprox 0.01$ and thus $\Delta\ltapprox 0.20\pm 0.01$.

Let us now consider the observable $O_2$. This quantity shows much 
larger corrections to scaling compared to the previous one: indeed in
no case we can identify a region of $N_{cut}$ where the estimates
are constant. In all cases the estimates of $\sigma$ are clearly increasing 
with $N_{cut}$. Using the results with $N_{cut}=8000$ we have
\be
\sigma \gtapprox \cases{0.991(5) & for geometry 1 \cr
                        0.980(6) & for geometry 2 \cr
                        0.986(6) & for geometry 3 \cr
                        0.964(7) & for geometry 4 \cr}
\ee
We conclude thus that $\sigma\gtapprox 0.99\pm 0.01$ so that 
$\Delta\ltapprox 0.19\pm 0.01$.

Consider now the observable $O_4$. It has much larger errors than
the two previous ones. Here we can only give a very rough estimate
$\sigma\approx 2.10(10)$ so that $\Delta\approx 0.25(10)$. 

Finally let us discuss the results for $\tilde{O}_2$. In this case the 
estimates of $\sigma$ are much lower than expected and indeed for 
both geometries 3 and 5 they seem to indicate, although without 
much confidence, $\sigma\approx 0.5\pm 0.2$. It thus appears that
this observable does not couple to the leading operator breaking the 
rotational invariance. This fact can be proved rigorously for the 
ordinary random walk: in Appendix A we show that for geometries 
3 and 5 we have indeed 
\be
\< \tilde{O}_2\>_N = 0
\ee
for all values of $N$. For the SAW $\< \tilde{O}_2\>_N \not=0$; however
our data show that for this observable, in formula \reff{coordinate}, 
not only $A$ but also $B_{\cal R}$ vanishes. Thus the study of 
$\< \tilde{O}_2\>_N$ allows to compute a new subleading exponent
$\Delta'>\Delta$: from $\sigma\sim 0.5$ we would get $\Delta'\sim 0.7$.
Notice however that the error bars are too big to really trust this estimate.
\begin{table}
\footnotesize
\begin{center}
\begin{tabular}{|c||c|c|c|c|}
\hline
geometry & cut &  $p_{\cal R} \pm \delta p_{\cal R}$ & $\Delta
\pm \delta \Delta $ & $CL $  \\
\hline \hline
 &   500 & $   0.182995 \pm   0.000067 $ &
 $   0.2150 \pm   0.0064 $ & 74.926\%\\
 \cline{2-5}
&  1000 & $   0.194002 \pm   0.000070 $ &
 $   0.215 \pm   0.012 $ & 58.808\%\\
 \cline{2-5}
 &  2000 & $   0.203434 \pm   0.000089 $ &
 $   0.223 \pm  0 .026 $ & 40.621\%\\
 \cline{2-5}
1 &  4000 & $   0.21173 \pm   0.00012 $ &
 $   0.16 \pm   0.13 $ & 56.695\%\\
 \cline{2-5}
 &  8000 & $   0.21861 \pm   0.00014 $ && \\
 \cline{2-5}
 & 16000 & $   0.22461 \pm   0.00027 $ &&\\
 \cline{2-5}
 & 32000 & $   0.23041 \pm   0.00058 $ &&\\
 \hline\hline
 &   500 & $   0.34624 \pm   0.00012 $ &
 $   0.2357 \pm   0.0067 $ & 83.318\%\\
 \cline{2-5}
 &  1000 & $   0.36415 \pm   0.00012 $ &
 $   0.234 \pm   0.013 $ & 69.572\%\\
 \cline{2-5}
 &  2000 & $   0.37928 \pm   0.00015 $ &
 $   0.237 \pm   0.027 $ & 48.991\%\\
 \cline{2-5}
2 &  4000 & $   0.39235 \pm   0.00019 $ &
 $   0.176 \pm   0.052 $ & 61.677\%\\
 \cline{2-5}
 &  8000 & $   0.40309 \pm   0.00022 $ &&\\
 \cline{2-5}
 & 16000 & $   0.41235 \pm   0.00044 $ &&\\
 \cline{2-5}
 & 32000 & $   0.42116 \pm   0.00094 $ &&\\
 \hline\hline
 &   500 & $   0.33456 \pm   0.00011 $ &
 $   0.2276 \pm   0.0063 $ & 74.457\%\\
 \cline{2-5}
 &  1000 & $   0.35209 \pm   0.00011 $ &
 $   0.226 \pm   0.012 $ & 58.934\%\\
 \cline{2-5}
 &  2000 & $   0.36696 \pm   0.00014 $ &
 $   0.232 \pm   0.026 $ & 39.799\%\\
 \cline{2-5}
 3&  4000 & $   0.37993 \pm   0.00018 $ &
 $   0.168 \pm   0.086 $ & 54.075\%\\
 \cline{2-5}
 &  8000 & $   0.39061 \pm   0.00021 $ &&\\
 \cline{2-5}
 & 16000 & $   0.39983 \pm   0.00041 $ &&\\
 \cline{2-5}
 & 32000 & $   0.40874 \pm   0.00088 $ &&\\
 \hline\hline
 &   500 & $   0.58302 \pm   0.00015 $ &
 $   0.2637 \pm   0.0067 $ & 75.629\%\\
 \cline{2-5}
 &  1000 & $   0.60509 \pm   0.00015 $ &
 $   0.259 \pm   0.013 $ & 64.062\%\\
 \cline{2-5}
 &  2000 & $   0.62332 \pm   0.00018 $ &
 $   0.258 \pm   0.027 $ & 43.116\%\\
 \cline{2-5}
 4&  4000 & $   0.63875 \pm   0.00023 $ &
 $   0.195 \pm   0.064 $ & 55.552\%\\
 \cline{2-5}
 &  8000 & $   0.65123 \pm  0 .00025 $ &&\\
 \cline{2-5}
 & 16000 & $   0.66182 \pm   0.00047 $ &&\\
 \cline{2-5}
 & 32000 & $   0.6718 \pm   0.0010 $ &&\\
 \hline\hline
 &   500 & $   0.46027 \pm   0.00013 $ &
 $   0.2397 \pm   0.0063 $ & 72.307\%\\
 \cline{2-5}
 &  1000 & $   0.48113 \pm   0.00013 $ &
 $   0.236 \pm   0.012 $ & 58.015\%\\
 \cline{2-5}
 &  2000 & $   0.49865 \pm   0.00016 $ &
 $   0.242 \pm   0.025 $ & 38.594\%\\
 \cline{2-5}
 5&  4000 & $   0.51380 \pm   0.00021 $ &
 $   0.177 \pm   0.072 $ & 53.900\%\\
 \cline{2-5}
 &  8000 & $   0.52619 \pm   0.00024 $ &&\\
 \cline{2-5}
 & 16000 & $   0.53683 \pm   0.00046 $ &&\\
 \cline{2-5}
 & 32000 & $   0.5470 \pm   0.0010 $ &&\\
 \hline\hline
 &   500 & $   0.74526 \pm   0.00016 $ &
 $   0.2897 \pm   0.0068 $ & 49.055\%\\
 \cline{2-5}
 &  1000 & $   0.76491 \pm   0.00014 $ &
 $   0.280 \pm   0.012 $ & 49.324\%\\
 \cline{2-5}
 &  2000 & $   0.78069 \pm   0.00017 $ &
 $   0.276 \pm   0.024 $ & 30.554\%\\
 \cline{2-5}
 6&  4000 & $   0.79389 \pm   0.00019 $ &
 $   0.206 \pm   0.052 $ & 57.904\%\\
 \cline{2-5}
 &  8000 & $   0.80432 \pm   0.00020 $ &&\\
 \cline{2-5}
 & 16000 & $   0.81312 \pm   0.00035 $ &&\\
 \cline{2-5}
 & 32000 & $   0.82126 \pm   0.00067 $ &&\\
\hline\hline
\end{tabular}
\end{center}
\caption{Probabilities of intersection $p_{\cal R}$ for the various 
geometries. Here $CL$ denotes the confidence level of the fit
used to determine $\Delta$.}
\label{tableprob}
\end{table}

Let us now discuss the behaviour of the intersection probabilities 
$p_{\cal R}$. As discussed before, pure dimensional arguments
suggest that the probability that a SAW intersects any one-dimensional 
set is vanishing in a three-dimensional space. This argument works
of course in the {\em continuum} limit. For SAWs on the lattice 
this statement translates in the fact that given a SAW of length $N$,
$\{\omega_i\}_{i=0,\ldots,N}$, and a one-dimensional set $S$ such that
$\omega_0$ is at a distance of order $N^\nu$ from $S$, then the 
probability that the walk intersects $S$ goes to zero for $N\to\infty$
as $N^{-\Delta}$. In our case $p_{\cal R}$ is however the probability
that the SAW intersects a one-dimensional set ${\cal R}$ which is at 
a finite fixed distance from the origin of the walk. In this case we expect 
the intersection probability to tend to a constant as $N\to\infty$ and 
thus a behaviour of the form
\be
p_{\cal R} (N) = p_{\cal R} (\infty) + {b_{\cal R}\over N^\Delta}
\label{pbehaviour}
\ee
The results for $p_{\cal R}$ for the various geometries are reported in
Table \ref{tableprob}, together with the estimates  of $\Delta$ 
from a three-parameter fit of the form \reff{pbehaviour}. The estimates 
indicate $\Delta=0.23(3)$ except for the geometry 6 in which case 
one would derive a much higher value of $\Delta$, $\Delta\approx 0.28$.
These discrepancies should not however be taken seriously: the fit 
\reff{pbehaviour} is very unstable in presence of additional 
subleading corrections. To understand the size of the systematic 
error one should expect from fits of the form \reff{pbehaviour}, we have 
performed the following test: we have considered $O_1$ and we have 
analyzed the data as 
\be
   {<O_1>_N\over N^\nu} = a + b N^{-\Delta}
\ee
where we have used $\nu = 0.5877$. In all cases we have obtained 
estimates $\Delta\approx 0.16-0.23$ and moreover we have found 
$a$ barely compatible with zero within error bars 
(for instance for the geometry 1, $N_{cut} = 4000$, we have
$a = (24 \pm 23)\cdot 10^{-4}$).
Clearly the additional corrections
play still an important role. It is however reassuring that the value of 
$\Delta$ is in essential agreement with what we expect.

\section{Conclusions}

In this paper we have studied the role played by one-dimensional 
vacancies in the critical behaviour of the three-dimensional SAW.
As already pointed out in \cite{Considine-Redner,CFP} a new critical 
exponent arises. We have carefully checked that the exponent 
depends only on the dimensionality of the vacancies by verifying its 
independence from the shape of the excluded region: in particular it 
does not depend on the discrete symmetry which the lattice has 
after the introduction of the excluded region.

We have given a geometrical argument to interpret the new exponent and we have
thus derived a classical prediction for it. Of course we expect 
renormalization effects to change the classical formula: we will 
thus write 
\be
\Delta = 2 \nu - 1 + \eta_{\cal R} \approx 0.175 + \eta_{\cal R}
\ee
The quantity $\eta_{\cal R}$ is an anomalous dimension which is 
expected to be small. Our numerical data give 
\be 
\Delta = 0.18 \pm 0.02 
\ee
so that $|\eta_{\cal R}|\ltapprox 0.02$. The classical prediction,
obtained setting $\eta_{\cal R}=0$ is thus a very good approximation.

Let us notice that our estimate of $\Delta$ is somewhat lower than
the estimate of \cite{Considine-Redner}, $\Delta\approx 0.24$ and 
of \cite{CFP}, $\Delta\approx 0.217\pm 0.013$. 
The origin of these discrepancies 
is probably in the neglected additional corrections to scaling: 
the results of \cite{Considine-Redner} are obtained from an 
exact enumeration and thus probe only very short walks, while the 
estimate of \cite{CFP} comes from walks which have mainly $N\approx 1000-4000$.
Here we use longer SAWs and a much higher statistics and thus we can 
do a much more detailed study of the role of the next subleading terms. 
We thus hope to have a better control of the 
additional corrections although it is conceivable that also our present
estimate is systematically higher than the true result. We hope that 
our error bar, which we believe is very conservative,
takes correctly into account these systematic effects.

Let us finally remark that all the arguments we have given are of extremely 
general nature and can thus be used for other systems and geometries.

\begin{appendix}

\section{Critical behaviour of random walks in presence of $d_{\cal R}$ 
dimensional vacancies}

Consider a $d$-dimensional lattice $\Omega$ and a
region ${\cal R}$.
Then let $c_N(x)$ be the number of ordinary random 
walks, starting from the origin and ending in $x$, that never intersect 
${\cal R}$ except for $t=0$. In this appendix we will derive 
a general integral equation for the generating function which
can be solved exactly when ${\cal R}$ is a $d_{\cal R}$-dimensional hyperplane.
In this way we will be able to check the computations of Section 2.

Let us start from the recursion relations
\bea
&&c_0\left({x}\right)=\delta\left({x},{0}\right)\;\; ,\nonumber\\
&&c_N\left({x}\right)  =  \sum^{d}_{i=1} \left[
c_{N-1}\left({x}+{e}_i \right)+c_{N-1}\left({x}-{e}_i
\right)\right]\cdot 
\left(1-\sum_{\alpha\in{\cal R}} \delta(x,\alpha) \right) \;\; 
\hbox{for } N\ge 1 \nonumber \\ [-2mm]
&& {} \label{recursion}
\eea 
where $e_i$ is the unit vector in the $i$-direction. Let us now introduce 
the generating function
\be
G( {x})= \sum^{\infty}_{N=0}\beta^N c_N({x})
\ee
and its Fourier transform $\hat{G}(q)$. It is then a simple matter to 
obtain the following equation
\bea
\hat{G}({q})&=&1 + \beta \,\hat{G}({q})\left(2d-{\hat q}^2\right)
\nonumber\\
&&   -\beta\int^{\pi}_{-\pi}
      \frac{d^dk}{(2\pi)^d} \,\hat{G}({k})
      \left(2d-{\hat k}^2\right)
      \sum_{\alpha\in {\cal R}} e^{i\left(k -q\right)\cdot {\alpha}}
\label{gf}
\eea
where $\hat q_i = 2\sin\left(\frac{q_i}{2}\right)$.
Define now the free propagator which coincides (apart from a factor
$\beta$) with the generating function for unconstrained random walks
\be
D(q) =\, \frac{1}{m_0^2+{\hat q}^2}
     =\, \frac{\beta}{1-2\beta\displaystyle\sum^{d}_{i=1} \cos q_i }
\label{propagator}
\ee
where $m_0^2=\frac{1-2\beta d}{\beta}$. Then \reff{gf} can be rewritten as 
\be
\hat{G}(q) = \, {1\over \beta} D(q) - D(q) 
      \int^\pi_{-\pi} {d^dk\over (2\pi)^d} \,\hat{G}(k) 
      (2 d - \hat{k}^2) \sum_{\alpha\in {\cal R}} 
       e^{i(k-q)\cdot\alpha}
\label{primasoluzione}
\ee
This equation is completely general and applies to any excluded region
${\cal R}$. 

Let us now restrict ourselves to the case in which ${\cal R}$ is 
the $d_{\cal R}$-dimensional
hyperplane  given by the equations $x_{d_{\cal R}+1}=\ldots=x_d=0$.
If we now multiply \reff{primasoluzione}
by $(2 d - \hat{q}^2) \sum_{\eta \in {\cal R}}
e^{i(q-k^{\prime})\cdot\eta}$ and integrate over $q$,
using the identity $2d-\hat{q}^2=\frac{1}{\beta}-{D(q)}^{-1}$,
we obtain 
\begin{eqnarray}
&& \int^\pi_{-\pi} {d^dq\over (2\pi)^d} \hat{G}(q) 
      (2 d - \hat{q}^2) \sum_{\alpha\in {\cal R}} 
       e^{i(q-k^{\prime})\cdot\alpha}\, = 
\nonumber \\
&& \qquad 
   {1\over \beta} \left[ 1 - \beta 
   \left(\int^\pi_{-\pi} {d^dq\over (2\pi)^d} D(q) 
      \sum_{\alpha\in {\cal R}} e^{i(q-k^{\prime})\cdot\alpha}\right)^{-1}
                  \right]
\end{eqnarray}
Inserting back in \reff{primasoluzione} we get finally
\be
\hat{G}(q) \, = \,
    {D(q)  \over \int^\pi_{-\pi} {d^{\cal D_R}k \over (2\pi)^{\cal
D_R}} D\left(q_1,\cdots,q_{d_{\cal R}},k_{{d_{\cal
R}}+1},\cdots,k_d\right)}
\label{integral}
\ee
where ${\cal D}_{\cal R}=d-d_{\cal R}$ .
It is easy to check that this solution has the correct properties. 
Indeed one can verify immediately that $G(x) = \delta_{x,0}$ whenever 
$x\in {\cal R}$. 
Notice finally that for $d_{\cal R}=0$ the solution 
does not become the standard solution for unconstrained random walks,
but gives the generating function for random walks with the excluded 
origin. 

Let us now consider the critical limit $m_0\to 0$ and let us define
\bea
I_{\cal R}(\hat{m}_0^2)  &\equiv& 
       \int^\pi_{-\pi} {d^{{\cal D}_{\cal R}}k \over 
	(2\pi)^{{\cal D}_{\cal R}}}
     D\left(q_1,\cdots,q_{d_{\cal R}},k_{d_{\cal R} + 1},\cdots,k_d\right)  =
\nonumber   \\
    && \int^\pi_{-\pi}  {d^{{\cal D}_{\cal R}}k \over 
	(2\pi)^{{\cal D}_{\cal R}}}
       {1 \over  {\hat m_0}^2 +{\hat k}^2}
\label{integral1}
\eea
where ${\hat m_0}^2 = m_0^2+{\hat{q}_{\|}}^2$ and ${{q}_{\|}}=\left(
q_1,\dots,q_{d_{\cal R}},0,\dots,0 \right)$.
The introduction of the excluded region will thus be relevant or irrelevant
depending on the behaviour of $I(\hat{m}_0^2)$ for $\hat{m}_0\to 0$. If the 
limit exists the perturbation is irrelevant, while if the integral diverges 
the perturbation is relevant or marginal. This last case corresponds to
${\cal D_R}=1,2$. 
For ${\cal D}_{\cal R}=1$, we have
\be
I_{\cal R}(\hat{m}_0^2) = \frac{1}{\hat m_0 \sqrt{{\hat m_0}^2+4}}
\ee
and the critical behaviour is drastically changed since
\be
\hat{G}(q)=\frac{\sqrt{m_0^2 +{\hat{q}_{\|}}^2} 
\sqrt{m_0^2+{\hat{q}_{\|}}^2+4}}{m_0^2+\hat{q}^2}\;\; .
\ee
In the critical limit we have 
\be
\hat{G}(q)= \frac{2 \sqrt{ m_0^2 +{q_{\|}}^2}
                  }{m_0^2+q^2} \;\; .
\ee
{}From this expression it is easy to verify that, for $N\to\infty$,
\be
c_N \equiv \sum_x c_N(x) \approx {2\over \sqrt{2 \pi d}} (2 d)^N N^{-1/2}
\ee
so that $\gamma=1/2$ which differs from the value of $\gamma$ for 
random walks in free space, $\gamma = 1$. Analogously we have
\bea
\< x^2_1 \>_N \approx {N\over d} \;\; ,\\
\< x^2_d \>_N \approx {2 N\over d} \;\; .
\eea
The exponent $\nu$ is not changed, but the amplitudes are, as expected,
dependent on the direction and different from the value they assume in
free space, $\< x^2_1 \>_N = \< x^2_d \>_N = N/d$.

If ${\cal D}_{\cal R}=2$, which, as we shall see, corresponds
to a marginal operator, we have 
\be
I_{\cal R}(\hat{m}_0^2)  =\, 
  \frac{2}{\pi}\frac{1}{{\hat m_0}^2+4} K\left(\frac{4}{{\hat m_0}^2+4}\right)
\ee
where $K(z)$ is an elliptic integral. 
For ${\hat m_0}^2 \to 0$, $I_{\cal R}(\hat{m}_0^2)$ 
diverges logarithmically so that we get in the critical limit
\be
\hat{G}(q) =\, 
 - {{4 \pi} \over (m_0^2 + q^2) \log[(m_0^2 +q_{\|}^2)/32] }
\ee
Thus the propagator differs from the 
unperturbed one only by a logarithmic correction.  
{}From this expression we easily get 
\be 
    c_N \equiv \sum_x c_N(x) \sim {(2 d)^N\over \log N} \;\; .
\ee
Thus in this case we have $\gamma=1$ as for random walks in free space:
however an additional logarithmic correction appears as expected in the
marginal case. Analogously we find
\be
\< x^2_1 - x^2_d \>_N \sim {N\over \log N} \;\; .
\ee
If now ${\cal D}_{\cal R} > 2$ $I_{\cal R}(\hat{m}_0^2)$ has a 
finite limit for $\hat{m}_0\to 0$ so that in the critical limit 
\be
\hat{G}(q) \, =\, {D(q)\over I_{\cal R}(0)}
\label{A14}
\ee
Thus $\hat{G}(q)$ is identical to the generating function of unconstrained
random walks except for a multiplicative constant which is related 
to the total number of walks  $c_N = \sum_x c_N(x)$. Indeed from \reff{A14}
it follows that for large $N$ 
\be 
c_N \approx {(2d)^N\over 2 d I_{\cal R}(0)}
\ee
so that $p_{{\cal R}}$, the probability that an unconstrained 
random walk intersects ${\cal R}$ is simply $1 - 1/(2 d I_{\cal R}(0))$.
It is easy to check that this probability tends to zero as $d\to\infty$
at $d_{\cal R}$ fixed. 
Indeed, let us compute the large-${\cal D}_{\cal R}$ expansion of 
$I_{\cal R}(0)$: we start from the standard representation 
of $I_{\cal R}(m_0^2)$ in terms of Bessel functions:
\be
I_{\cal R}(\hat{m}_0^2) = 
\, \int^\infty_0 {dt\, e^{-t\left(\hat{m}_0^2+2{\cal D_R}\right)}}
{I_0 ^{\cal D_R}(2t)}
\label{irrel}
\ee
where $I_0 (t)$ is the zeroth-order Bessel function of first kind.
Expanding $I_0(t)$ around $t=0$ we get finally
\bea
I_{\cal R}(\hat{m}_0^2)& = &
 \frac{1}{{\hat m_0}^2 + 2{\cal D_R} } \, \times
\nonumber \\
& &
\sum_{n_1=0}^\infty \cdots \sum_{n_{\cal D_R}=0}^\infty
{ (2n_1+\cdots+2 n_{\cal D_R})! \over (n_1!\cdots n_{\cal D_R}!)^2}
{ 1 \over ( {\hat m_0}^2 + 2 {\cal D_R})^{2(n_1+\cdots+n_{\cal D_R})}}
\eea
so that, for $ {\cal D}_{\cal R} \to \infty $,
\be
I_{\cal R}(0) \approx {1\over 2{\cal D}_{\cal R}} \left( 
   1 + {1\over 2{\cal D}_{\cal R}} + 
     {3\over 4{\cal D}^2_{\cal R}} + \ldots \right)
\ee
It follows that for $d\to\infty$, 
$p_{\cal R} \approx (2 d_{\cal R} + 1)/(2 d)$.

Let us now discuss the subleading corrections to \reff{A14}.
We need here the small-$\hat{m}_0^2$ expansion of $I(\hat{m}_0^2)$.
We start from the well-known
asymptotic expansion for large $t$ of the Bessel function ${I_0}(2t)$
\be
I_0^{\cal D_R}(2t) \approx e^{2{\cal D_R}t} \sum_{n=0}^{\infty}
\frac{b_n({\cal D_R})}{(2t)^{\frac{{\cal D_R}}{2} + n} } \;\; .
\ee
Using \reff{irrel} we can rewrite our integral as 
\bea
I_{\cal R}(\hat {m}_0^2)&\approx& \int_0^{1}{dt\, e^{-t{\hat {m}_0^2}}
\left[e^{-2t}{I_0}(2t)\right]^{\cal D_R}} + 
\nonumber \\
& &
\int_1^{\infty}{dt\, e^{-t{\hat {m}_0^2}}
\left[\left(e^{-2t}{I_0}(2t)\right)^{\cal D_R} -\sum_{n=0}^{\infty}
{\frac{b_n({\cal D_R})}{(2t)^{\frac{{\cal D_R}}{2} + n} } }\right] }+
\nonumber \\
& &
\int_1^{\infty}{dt\, e^{-t{\hat {m}_0^2}}}\sum_{n=0}^{\infty}
{\frac{b_n({\cal D_R})}{(2t)^{{\cal D_R}/2 + n} } }
\eea
The first two integrals have a regular expansion in terms of powers 
of $\hat{m}_0^2$ .

Let us now determine the behaviour in $\hat{m}_0^2$ of the generic
term appearing in the series in the last term. Integrating by parts
we get
\bea
\lefteqn{
\int_1^\infty
{dt \, e^{-t{\hat{m}_0^2}}\frac{b_n({\cal D_R})}{(2t)^{{\cal D_R}/2 + n}
 } } =
\frac{ \hat{m}_0^{{\cal D_R} + 2 n -2}}{2^{{\cal D_R}/2 + 
n}} \,b_n({\cal D_R})
\int_{\hat {m_0}^2}^{\infty}
{d\xi \frac{e^{-\xi}}{\xi^{{{\cal D}_{\cal R}/2 + n}}}}=    } 
  \hspace{2cm} \nonumber  
\\ 
&& \frac{b_n ({\cal D}_{\cal R})}{2^{{\cal D_R}/2 + n}} 
\left\{
   \left[\sum_{l=1}^{g(k,n)} \hat{m}_0^{2l-2} (-1)^{l-1} 
       \left(\prod_{i=1}^l {1 \over {\cal D_R}/2 + n-i}\right) 
   \right]\
   \left[\sum_{s=0}^\infty \frac{ (-1)^s}{s!} \hat{m}_0^{2s}\right]
+ \right.
\nonumber\\
&&
\left. (-1)^{g(k,n)} \hat{m}_0^{{\cal D_R} + 2 n - 2} F(\hat{m}_0^2)
\left(\prod_{i=1}^{g(k,n)} {1 \over {\cal D_R}/2 + n-i}\right)\right\}
\eea
where, for $k$ integer,
\bea
\hskip -1.2truecm
g(k,n) &=& \cases{k+n-1 & if ${\cal D_R}=2k$; \cr
 k+n+1 &  if ${\cal D_R}=2k+1$\cr}   \\
\hskip -1.2truecm
F(\hat{m}_0^2) &=& 
 \cases{- \hbox{\rm Ei} (- \hat{m}_0^2)
     = \, -\log \hat{m}_0^2 - \gamma_E
      -\sum_{k=1}^{\infty}\frac{(-1)^{k}{\hat{m}_0^{2k}}}{k\cdot k!} 
                    & if ${\cal D_R}=2k$; \cr
       {\sqrt{\pi}\over2} - 
            \hat{m}_0^3 \sum_{k=0}^{\infty}
            \frac{(-1)^{k}{\hat{m}_0^{2k}}}{(k+{3\over 2})\cdot k!}
                    &  if ${\cal D_R}=2k+1$\cr}
\eea
Here $\gamma_E$ is the Euler-Mascheroni constant, 
$\gamma_E \approx 0.5772156649$.
The whole integral can be represented in terms of
$\hat{m}_0^2$ by
\be
I_{\cal R}(\hat {m}_0^2)=\sum_{n=0}^{\infty}A_n \,\hat {m}_0^{2 n} +
   \hat {m}_0^{ {\cal D_R} - 2} 
  C(\hat{m}_0^2) \;\; ;
\label{IRasy}
\ee
$A_n$ are suitable constants and 
$C(\hat{m}_0^2)$ is a function of $\hat{m}_0^2$ which is 
finite for $\hat{m}_0^2\to0$ for ${\cal D_R}$ odd,
diverging logarithmically for ${\cal D_R}$ even. The
second term in \reff{IRasy}
represents the effect of the excluded 
region and corresponds to an exponent $\Delta$
\be
\Delta = \nu({\cal D_R} - 2) = {1\over2}{\cal D_R}  - 1
\ee
which agrees with our prediction.

To conclude this appendix let us prove a result for more general 
excluded regions which we will use in the main text. Let the 
excluded region be of the form ${\cal R} = \cup_{i=1}^d {\cal R}_i$
where ${\cal R}_i$ is some subset of points of the $i$-th coordinate axis.
If $i_1\not= i_2 \ldots \not= i_n$, $2\le n \le d$, then
\be
\< x_{i_1} x_{i_2} \ldots x_{i_n} \>_N =\, 0
\label{smallth}
\ee
In particular, for geometries 3 and 5 we have $\<{\tilde O}_2\>_N=0$.

To prove \reff{smallth} consider \reff{primasoluzione}, which, as 
we already said, is valid for general excluded regions $\cal R$. Using now
\bea
\left. {\partial\over \partial q_{i_1}}
{\partial\over \partial q_{i_2}} \ldots
{\partial\over \partial q_{i_n}} D(q) \right|_{q=0} &=& 0 \\
{\partial\over \partial q_{i_1}}
{\partial\over \partial q_{i_2}} \ldots
{\partial\over \partial q_{i_n}} 
\sum_{\alpha\in {\cal R}} e^{i(k-q)\cdot\alpha} &=& 0
\eea
for $i_1\not= i_2 \ldots \not= i_n$, $2\le n \le d$, we get
\be
\left. {\partial\over \partial q_{i_1}}
{\partial\over \partial q_{i_2}} \ldots
{\partial\over \partial q_{i_n}} \hat{G}(q) \right|_{q=0} =\, 0
\ee
from which \reff{smallth} immediately follows.

\begin{table}
\begin{center}
\begin{tabular}{|c||c|c|c|}
\hline
$ N $ & $\< x^2 + y^2 + z^2\>_N$ &
        $\< x^4 + y^4 + z^4\>_N$ & $Q_N$ \\
\hline
 $ 500 $ & $     1785.8 \pm     1.1 $ & 
           $    (28727 \pm    38)\cdot 10^2 $ & $ 
            0.90078 \pm   0.00046 $\\
 $1000 $ & $     4051.8 \pm     2.0 $ & 
           $    (14826 \pm    16)\cdot 10^3 $ & $ 
            0.90309 \pm   0.00039 $\\
 $2000 $ & $     9167.4 \pm     4.4 $ & 
          $    (76027 \pm    81)\cdot 10^3 $ & $ 
            0.90464 \pm   0.00038 $\\
 $4000 $ & $    20752.9 \pm     9.4 $ & 
          $    (38994 \pm    39)\cdot 10^4 $ & $ 
            0.90539 \pm   0.00036 $\\
 $8000 $ & $    46995 \pm    18 $ & 
           $    (20020 \pm    17)\cdot 10^5 $ & $ 
            0.90649 \pm   0.00032 $\\
 $16000 $ & $   106166 \pm    60 $ & 
           $    (10222 \pm    13)\cdot 10^6 $ & $ 
            0.90692 \pm   0.00047 $\\
 $32000 $ & $   240356 \pm   221 $ & 
           $    (5247 \pm    11)\cdot 10^7 $ & $ 
            0.90829 \pm   0.00078 $\\
\hline
\end{tabular}
\end{center}
\caption{Mean values for SAWs in absence of any excluded region.
$(x,y,z)$ are the coordinates of the end-point of the walk.}
\label{tabellasenzaneedles}
\end{table}

\section{Large-distance behaviour of the two-point function}

In this Appendix we will present some results which concern the 
three-dimensional self-avoiding walk with no excluded region and 
we will use it to discuss, along the lines of 
\cite{Fisher-Aharony,CPRV},
the behaviour of the two-point function $G(\vec{r};\beta)$ in the 
large-distance region $|r|\gtapprox R_e(\beta)$ where 
$R_e(\beta)$ is the mean end-to-end distance. 
Consider now the Fourier transform $\hat{G}(p;\beta)$; for 
$\beta\to\beta_c = 1/\mu$ standard scaling theory predicts 
\be
{\hat{G}(p;\beta)\over \hat{G}(0;\beta)} =\, \tilde{G}(q) 
\ee
where $q = p R_e(\beta)/6$. An important characteristic of 
$\tilde{G}(q)$ is the fact that in the region $q^2 \ltapprox 1$, 
$\tilde{G}(q)$ is essentially a free-field propagator, 
i.e. it can be parametrized as
\be
\tilde{G}(q) \approx {1\over 1 + q^2} \;\; .
\ee
The deviations are small and can be parametrized by a $(q^2)^2$ term, i.e.
by 
\be
\tilde{G}(q) \approx {1\over 1 + q^2 + b_2 (q^2)^2} \;\; .
\ee
A strong-coupling (exact-enumeration) study \cite{CPRV} set a bound on 
$b_2$:
\be 
  -3\cdot 10^{-4} \ltapprox b_2 \ltapprox 0\;\; .
\label{b2sc}
\ee
The constant $b_2$ has also been computed \cite{CPRV} 
in the $\epsilon$-expansion
and in the expansion in fixed dimension with the result : 
$b_2= -3 \cdot 10^{-4}$.

Here we want to give a bound on $b_2$ using our Monte Carlo data. A simple 
computation gives
\be
b_2 = 1 - {1\over 2} 
     {\Gamma(\gamma)\Gamma(\gamma+4\nu)\over \Gamma(\gamma+2\nu)^2}\, Q
\label{b2daQ}
\ee
where
\be
Q = \lim_{N\to\infty} Q_N \equiv 
    \lim_{N\to\infty} 
    {\< x^4 + y^4 + z^4\>_N\over \< x^2 + y^2 + z^2\>_N^2} \;\; .
\ee
Our Monte Carlo estimates for $Q_N$ are reported in Table
\ref{tabellasenzaneedles}. It is evident that the data show strong corrections
to scaling. To determine $Q$ we have thus performed a fit of the 
form
\be
Q_N = Q + {A\over N^\Delta} \;\; .
\ee
We find 
\begin{eqnarray}
Q &=& 0.9091 \pm 0.0016 \;\; ,\\
\Delta &=& 0.41 \pm 0.16 \;\; ,\\
\chi^2 &=& 1.77 \;\;\;\;\; (4\ {\hbox{\rm d.o.f.}}) \;\; .
\end{eqnarray}
The value of $\Delta$ is in agreement with the estimates of 
\cite{Sokal2}. Using for $\gamma$ the value \cite{gammaprep}
$\gamma=1.1575(5)$ we get finally
\be
b_2 = - (13 \pm 17) \ 10^{-4} \;\; .
\ee
Our Monte Carlo data confirm the fact that $b_2$ is extremely small
although we are unable to compute the actual value. 

On the other hand we can use \reff{b2sc} and \reff{b2daQ} together 
with the estimates of $\gamma$ and $\nu$ to obtain an estimate of $Q$.
We get $Q = 0.9082(11)$.

\end{appendix}

\end{document}